\documentclass[prl,twocolumn,citeautoscript,amsmath,superscriptaddress]{revtex4-1}

\usepackage{graphicx}
\usepackage{newtxtext,newtxmath}
\usepackage{epsfig}
\usepackage{amsmath,bbm,amssymb,bm,times}
\usepackage[colorlinks,linkcolor=blue,citecolor=blue]{hyperref}

\newcommand{\eqn}[1]{
\begin{eqnarray}
	#1
\end{eqnarray}
}

\begin{document}

\title{Exceptional non-Hermitian topological edge mode and its application to active matter}

\author{Kazuki Sone}
\email{sone@noneq.t.u-tokyo.ac.jp}
\affiliation{Department of Applied Physics, The University of Tokyo, 7-3-1 Hongo, Bunkyo-ku, Tokyo 113-8656, Japan}
\author{Yuto Ashida}
\affiliation{Department of Applied Physics, The University of Tokyo, 7-3-1 Hongo, Bunkyo-ku, Tokyo 113-8656, Japan}
\author{Takahiro Sagawa}
\affiliation{Department of Applied Physics, The University of Tokyo, 7-3-1 Hongo, Bunkyo-ku, Tokyo 113-8656, Japan}
\affiliation{Quantum-Phase Electronics Center (QPEC), The University of Tokyo, 7-3-1 Hongo, Bunkyo-ku, Tokyo 113-8656, Japan}

\begin{abstract}
Topological materials exhibit edge-localized scattering-free modes protected by their nontrivial bulk topology through the bulk-edge correspondence in Hermitian systems. While topological phenomena have recently been much investigated in non-Hermitian systems with dissipations and injections, the fundamental principle of their edge modes has not fully been established. Here, we reveal that in non-Hermitian systems robust gapless edge modes can ubiquitously appear owing to a mechanism that is distinct from bulk topology, thus indicating the breakdown of the bulk-edge correspondence. The robustness of these edge modes originates from yet another topological structure accompanying the branchpoint singularity around an exceptional point, at which eigenvectors coalesce and the Hamiltonian becomes nondiagonalizable. Their characteristic complex eigenenergy spectra are applicable to realize lasing wave packets that propagate along the edge of the sample. We numerically confirm the emergence and the robustness of the proposed edge modes in the prototypical models. Furthermore, we show that these edge modes appear in a model of chiral active matter based on the hydrodynamic description, demonstrating that active matter can exhibit an inherently non-Hermitian topological feature. The proposed general mechanism would serve as an alternative designing principle to realize scattering-free edge current in non-Hermitian devices, going beyond the existing frameworks of non-Hermitian topological phases.
\end{abstract}
\maketitle
Since topological materials exhibit robust scattering-free currents along the edges of samples, the notion of topology has attracted much interest both in fundamental physics and in device engineering. The first discovery of such a topological material is the integer quantum Hall effect \cite{Klitzing1980}, where it has been established that the gapless edge modes exactly correspond to the bulk topological number \cite{Thouless1982}. This has revealed a fundamental principle called the \textit{bulk-edge correspondence}. Nowadays, the topological materials have been found in much broader situations especially in the presence of symmetry, such as the time-reversal symmetry in topological insulators \cite{Kane2005,Hasan2010}. In such situations, the bulk-edge correspondence is still valid and predicts the presence or absence of robust edge modes \cite{Hasan2010}, from which a periodic table has been obtained \cite{Kitaev2009}. These discoveries have opened up a stream of material designs on the basis of the bulk band topology \cite{Hasan2010,Ando2013}. 

Although the conventional notion of topological materials is based on Hermitian Hamiltonians, effective Hamiltonians can become non-Hermitian in nonconservative systems including both quantum and classical ones, such as photonics \cite{Ruschhaupt2005,El-Ganainy2007,Ruter2010,Zhao2019}, ultracold atoms \cite{Ashida2017,Li2019}, optomechanics \cite{Xu2016,Jing2017}, electronic circuits \cite{Schindler2011,Ezawa2019}, mechanical lattices \cite{Kane2013,Nash2015}, and biophysical systems \cite{Murugan2017}. For example, in photonic systems non-Hermiticity can be introduced by engineering optical gain and loss through semiconductor amplifiers and acoustic modulators. Classification of non-Hermitian topological materials has been explored in terms of bulk band topology \cite{Zhao2019,Hu2011,Esaki2011,Leykam2017,Shen2018,Gong2018,Kunst2018,Xiong2018,Yao2018,Lee2019,Yokomizo2019,Zhou2019,Kawabata2019,Borgnia2020}, and a periodic table has been proposed in the same spirit as in the Hermitian case \cite{Gong2018,Zhou2019,Kawabata2019}. However, the bulk-edge correspondence is more subtle in non-Hermitian systems than in Hermitian systems, as unexplored non-Hermitian effects may protect unpredicted edge modes or remove edge modes from topologically nontrivial systems.

In this Article, we reveal a ubiquitous mechanism for realizing robust gapless edge modes, which emerge independently of the bulk topology and instead are protected in an unconventional manner unique to non-Hermitian systems. This indicates that the bulk-based classification cannot conclusively predict the existence or absence of edge modes in the non-Hermitian case, thus implying the breakdown of the bulk-edge correspondence. In the conventional topologically nontrivial systems, the proposed mechanism can further stabilize the gapless edge modes, even against symmetry-breaking disorder. These edge modes inherently exhibit large positive imaginary parts of the eigenenergies and thus are naturally applicable to topological insulator laser \cite{Harari2018,Bandres2018,Song2019}, where the amplified unidirectional wave packet propagates along the edge of the sample. We demonstrate the emergence of the proposed gapless edge modes and the lasing wave packets by performing numerical calculations of the prototypical tight-binding models.

\begin{figure*}
  \includegraphics[width=140mm,bb=0 0 1060 610,clip]{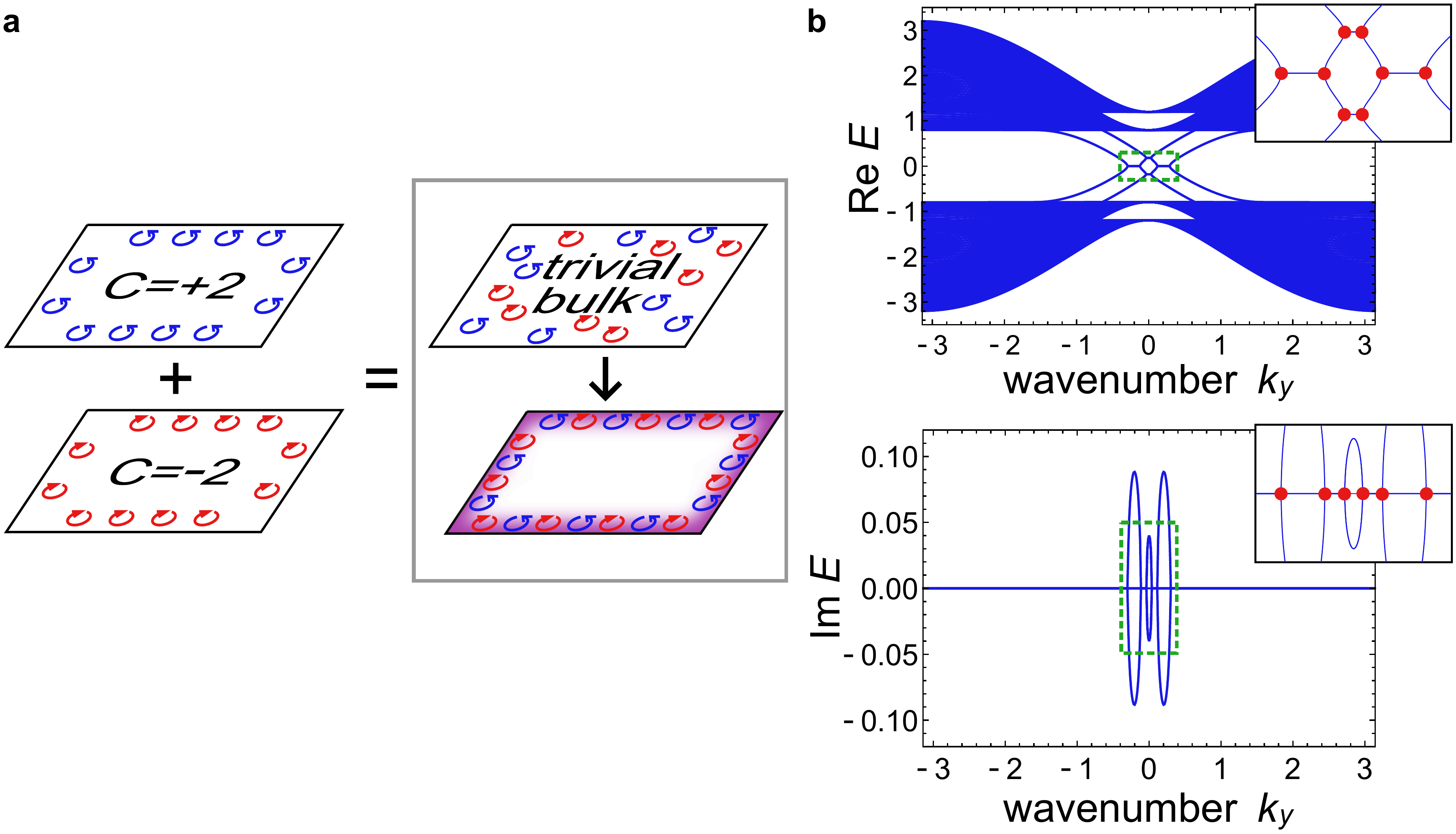}
\caption{\label{fig1}
Schematic figure and prototypical band structure of exceptional edge modes. (a) The system is composed of two Chern insulators which have the Chern numbers with the same absolute values and the opposite signs. If we consider Chern insulators with even Chern numbers, the bulk of the combined system becomes topologically trivial. However, when we use a non-Hermitian coupling to combine two Chern insulators, robust gapless modes can appear at the edge of the sample. (b) Two-layered non-Hermitian Bernevig-Hughes-Zhang model is considered to demonstrate the prototypical band structure of exceptional edge modes. We numerically calculate the edge band structure of the $1\times50$ ribbon-shaped system under the open boundary condition in the $x$ direction and the periodic boundary condition in the $y$ direction. The parameters used are $u=-1$, $c=0.2$, $\beta=0.14$, $\beta'=0.06$, and $\gamma=0.05$. Four gapless bands per edge exist in the bulk energy gap (they are doubly degenerated) and thus imply the topologically trivial bulk of the system. In the edge bands, we also find exceptional points (EPs) (indicated by red points in the insets), which are the wavenumbers at which the edge eigenstates coalesce and the Hamiltonian becomes nondiagonalizable. We can confirm that the imaginary parts of the eigenenergies appear from the EPs. These EPs play the role of the "glue" of the edge modes and thus prevent gap opening by perturbations or disorders, which is shown in Fig.~\ref{fig3}.
}
\end{figure*}

Our edge modes owe their robustness to the distinct topological structure of exceptional points (EPs), and thus here we term these modes as \textit{exceptional edge modes}. The EP \cite{Kato1966} is a singular point in the parameter space at which two or more eigenvectors and eigenvalues coalesce and a parameterized Hamiltonian becomes nondiagonalizable. The EP is unique to non-Hermitian systems and induces intriguing phenomena, such as interchanging eigenvectors after encircling an EP \cite{Dembowski2001}, coherent perfect absorption \cite{Longhi2010}, and unidirectional invisibility \cite{Lin2011}. The existence of EPs is supported by the nontrivial topology of the branchpoint singularity in intersecting Riemann surfaces around them \cite{Persson2000,Dembowski2001}. In one-dimensional systems like the edge modes of two-dimensional bulk systems, the emergence of EPs can be guaranteed by satisfying certain symmetries, such as the $PT$ symmetry, the $CP$ symmetry, the pseudo-Hermiticity, and the chiral symmetry \cite{Okugawa2019,Budich2019,Kawabata2019b}. EPs can disappear if either the symmetry is broken or a pair of EPs coalesce; the latter is reminiscent of the pair-annihilation of Weyl points \cite{Matsuura2013,Armitage2018} in Hermitian systems. We discover a general mechanism that EPs join two edge dispersions like glue and make them robust against disorder, which cannot be predicted by the existing periodic tables of topological phases \cite{Gong2018,Zhou2019,Kawabata2019}.

Furthermore, we explicitly show the existence of exceptional edge modes in a more realistic system based on active matter \cite{Marchetti2013}, which is a collection of self-propelled particles and has recently attracted much interest as a useful platform to study biological and out-of-equilibrium physics. Recent studies \cite{Souslov2017,Shankar2017,Dasbiswas2018,Souslov2019,Sone2019,Yang2019} have shown the existence of the edge modes in active matter protected by the bulk topology. Some of them \cite{Dasbiswas2018,Souslov2019,Yang2019} have utilized chiral active matter, which moves in a circular path or self-rotates. Chiral active matter has been experimentally realized, for example, in bacteria \cite{DiLuzio2005} and artificial L-shaped particles \cite{Kummel2013}. The hydrodynamics \cite{Furthauer2012,Banerjee2017} and the phase separation \cite{Ai2018} of chiral active matter have also been analyzed in recent studies. The effective Hamiltonian of the linearized hydrodynamic equations in active matter is, in general, non-Hermitian because of inherent dissipations and energy injections therein. We demonstrate that this type of non-Hermitian chiral active matter provides an ideal platform to experimentally realize the proposed exceptional edge modes.

\begin{figure*}
  \includegraphics[width=140mm,bb=0 0 1700 920,clip]{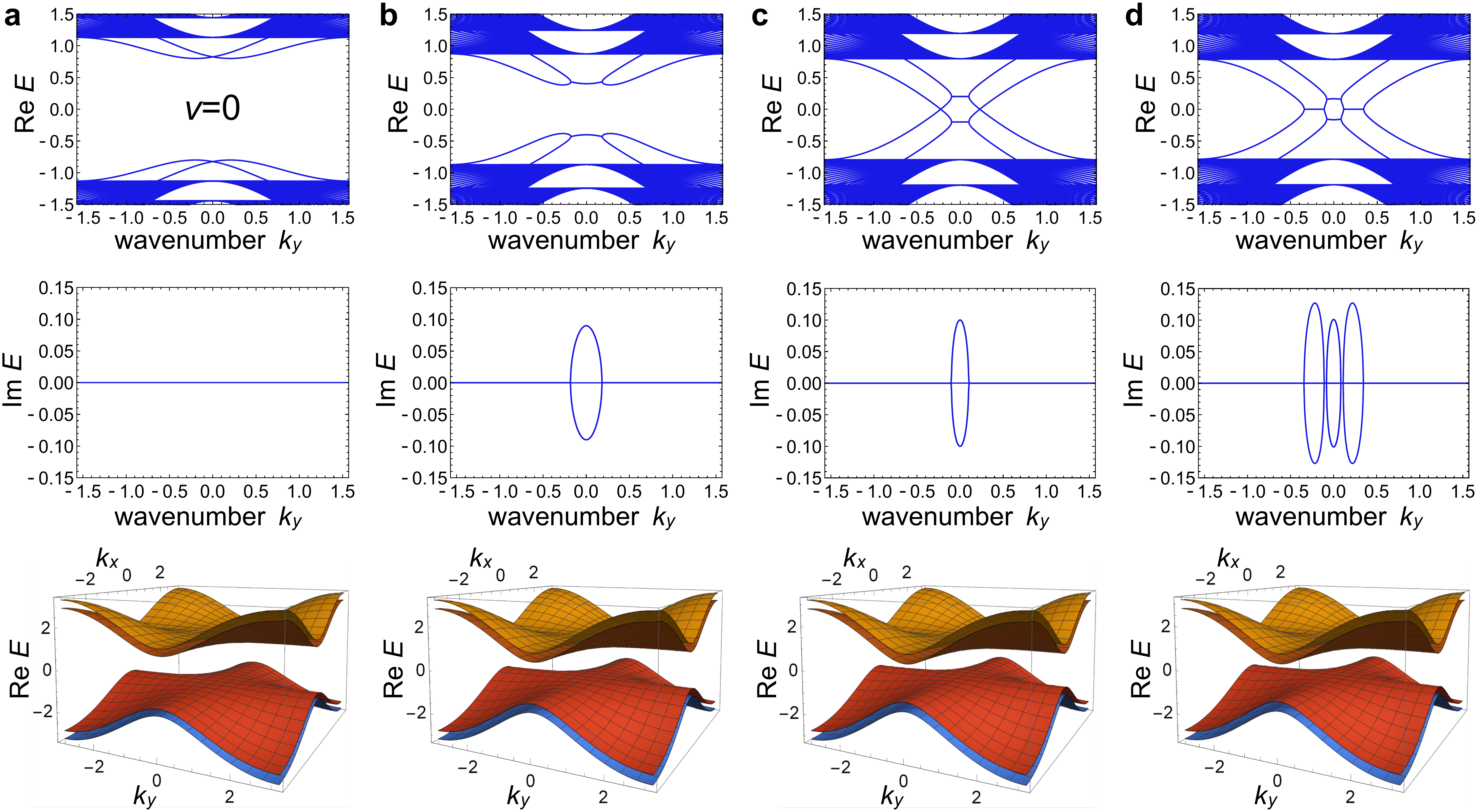}
\caption{\label{fig2}
Emergent exceptional edge modes through a non-Hermitian coupling. (a)-(d) Non-Hermiticity is increased from (a) to (d). The upper two figures show the real and imaginary parts of the edge dispersions. The lowest figure represents the bulk band structure of the homogeneous system under the periodic boundary conditions. We set the parameters $u=-1$, $c=0.2$, and $\alpha=0.8$ throughout the calculations in this Figure. (a) Without non-Hermitian coupling ($\beta=\beta'=0$), there is a large energy gap even in the edge band structure. Therefore, the bulk topology is trivial in this Hermitian system. (b) With increased non-Hermiticity ($\beta=0.81$, $\beta'=0.63$), the upper and lower edge dispersions approach to each other. (c) At the critical strength of the non-Hermitian coupling ($\beta=0.9$, $\beta'=0.7$), the upper and lower edge dispersions coalesce. However, there are no gap closings in the bulk band structure. Thus, the bulk topology should remain trivial. (d) With a stronger non-Hermitian coupling than the critical value ($\beta=0.909$, $\beta'=0.707$), pairs of EPs appear. These edge modes are robust against disorder and thus exceptional edge modes robustly appear even with trivial bulk topology.}
\end{figure*}

{\it Exceptional edge modes in two-layered non-Hermitian Bernevig-Hughes-Zhang model.---}
We first construct and analyze a minimal tight-binding model. For a Hermitian Hamiltonian $H$, time-reversal symmetry means that there exists a unitary operator $T$ satisfying $TH(\mathbf{k})T^{-1} = H^{\ast}(-\mathbf{k})$, where $H(\mathbf{k})$ is the Bloch Hamiltonian constructed from $H$ \cite{Hasan2010}. The definition of time-reversal symmetry can be extended to non-Hermitian systems and it has been pointed out \cite{Esaki2011,Gong2018,Zhou2019,Kawabata2019} that there are two types of time-reversal symmetry, i.e., $TH(\mathbf{k})T^{-1} = H^{\ast}(-\mathbf{k})$ and $TH(\mathbf{k})T^{-1} = H^{T}(-\mathbf{k})$, which are equivalent in Hermitian systems while not in non-Hermitian cases. One can construct the conventional time-reversal-symmetric topological insulator by coupling a Chern insulator with its time-reversal counterpart \cite{Kane2005,Bernevig2006}. The bulk bands of a time-reversal-symmetric insulator are topologically characterized by the $\mathbb{Z}_2$ index \cite{Kane2005,Hasan2010}, which corresponds to the parity of the number of the edge modes across the Fermi energy. If we construct a time-reversal-symmetric system from Chern insulators with even numbers of edge modes, we obtain a topologically trivial bulk. However, we reveal that such a trivial bulk can still accompany robust gapless edge modes by introducing non-Hermitian coupling between the two Chern insulators (see Fig.~\ref{fig1}a).

To construct the minimal model for demonstrating the emergence of such edge modes, we consider the two-layered Qi-Wu-Zhang (QWZ) model, $H_0 = I_2 \otimes H_{\rm QWZ} + c \sigma_x \otimes I_2$, which exhibits two chiral modes per edge in the bulk energy gap. Here, $H_{\rm QWZ}$ is the Hamiltonian of the QWZ model \cite{Qi2006}, which can be described as $H_{\rm QWZ}(\mathbf{k}) = \sin k_x \sigma_x + \sin k_y \sigma_y + (u+\cos k_x + \cos k_y) \sigma_z$ in the wavenumber space (see Supplementary Information for the real-space description). Here, $I_2$ is the $2\times2$ identity matrix and $\sigma_i$ is the $i$th component of the Pauli matrices. Also, we assume that $c$ is real, and thus the Hamiltonian is still Hermitian. By coupling $H_0$ and its time-reversal counterpart $H_0^{\ast}$ with a non-Hermitian term, $i\Sigma = i(\beta+\beta')I_2\otimes\sigma_x/2+i(\beta-\beta')\sigma_z\otimes\sigma_x/2$ with $\beta$, $\beta'$ being real parameters, we obtain the following non-Hermitian Hamiltonian
\begin{equation}
 H = \left(
  \begin{array}{cc}
   H_0 & i\Sigma \\
   i\Sigma & H^{\ast}_0 
  \end{array}
  \right) \label{toy-model1}.
\end{equation}
We note that this model resembles the Bernevig-Hughes-Zhang model \cite{Bernevig2006} but differs from it since our model has two layers of the QWZ model and two other layers of the time-reversal QWZ model which are coupled by the non-Hermitian term. Furthermore, since this Hamiltonian has the pseudo-Hermiticity defined as $\eta H(\mathbf{k})\eta^{-1}=H^{\dagger}(\mathbf{k})$ which can lead to another topological classification characterized by the $\mathbb{Z}$ invariant \cite{Zhou2019,Kawabata2019}, we add a Hermitian coupling and consider the Hamiltonian $H' = H + \gamma \sigma_x \otimes \sigma_y \otimes \sigma_x$,  to break the pseudo-Hermiticity. The additional Hermitian coupling corresponds to the spin coupling in condensed matter and thus can open an energy gap in the conventional trivial insulator. We can confirm that this Hamiltonian has time-reversal symmetry $TH'(\mathbf{k})T^{-1} = H'^{\ast}(-\mathbf{k})$, and thus have to consider $\mathbb{Z}_2$ indices as in Hermitian systems (see Supplementary Information). Below we focus on the parameter regimes in which the bulk bands are trivial in the conventional sense, i.e., the number of the edge modes in $H_0$ is even.

To reveal the existence of robust edge modes, we calculate the band structure of our model with open (periodic) boundaries in the $x$ ($y$) direction. Figure \ref{fig1}b shows the band structure for the wavenumber in the $y$ direction. There, gapless edge bands exist in the bulk energy gap and they accompany EPs, where both the eigenenergies and the eigenstates coalesce. The EPs act as a "glue" that holds the edge band structures together and thus stabilize the existence of exceptional edge modes. This gluing is reminiscent of the branchpoint structure in non-Hermitian bulk bands \cite{Zhou2018,Budich2019}, which remains until the EPs coalesce. In general models including the present one, the edge modes between two EPs exhibit the large imaginary parts of the eigenenergies, while all the bulk modes can have zero imaginary parts of the eigenenergies. As discussed below, this property finds a possible application to realize a topological insulator laser \cite{Harari2018,Bandres2018}.

To explicitly demonstrate that the appearance of EPs is independent of the bulk topology and thus violates the bulk-edge correspondence, we numerically calculate the edge band structures for different strengths of the non-Hermitian coupling. By modifying the strengths of the non-Hermitian coupling $\beta$, $\beta'$, we can control the existence of edge modes and EPs in the bulk gap as shown in Fig.~\ref{fig2}. On the other hand, during this modification, the bulk energy gap remains open. Therefore, the bulk topology should remain trivial for arbitrary strength of the non-Hermitian coupling and thus have no relation to the exceptional edge modes. This result indicates that while the bulk band topology still can predict the existence of the ordinary edge modes without exceptional points, it fails to predict the existence of robust exceptional edge modes.

While we have concentrated on time-reversal-symmetric systems so far, time-reversal symmetry is not the prerequisite for realizing exceptional edge modes. If the sum of the Chern numbers of the bulk bands below the energy gap is zero, a system without relevant symmetries cannot exhibit gapless edge modes protected by bulk topology. However, combining the topological systems with the opposite Chern numbers by the non-Hermitian coupling, we can obtain not only a trivial bulk but also robust exceptional edge modes. 

We can also realize exceptional edge modes in topologically nontrivial systems. We construct the non-Hermitian Bernevig-Hughes-Zhang model,
\begin{equation}
 H = \left(
  \begin{array}{cc}
   H_{\rm QWZ} & i\Sigma' \\
   i\Sigma' & H^{\ast}_{\rm QWZ} 
  \end{array}
  \right), \label{toy-model-nontrivial}
\end{equation}
where $i\Sigma'$ is the non-Hermitian coupling $i\Sigma'=i\beta \sigma_x$, and $\beta$ is real. This model satisfies the time-reversal symmetry and associates with a nontrivial $\mathbb{Z}_2$ invariant. We calculate the edge band structure and confirm the existence of the exceptional edge modes (see Supplementary Information). We note that the exceptional edge modes can also exist robustly against time-reversal-symmetry-breaking disorder and thus can be more advantageous than conventional edge modes.

\begin{figure*}
  \includegraphics[width=140mm,bb=0 0 1095 730,clip]{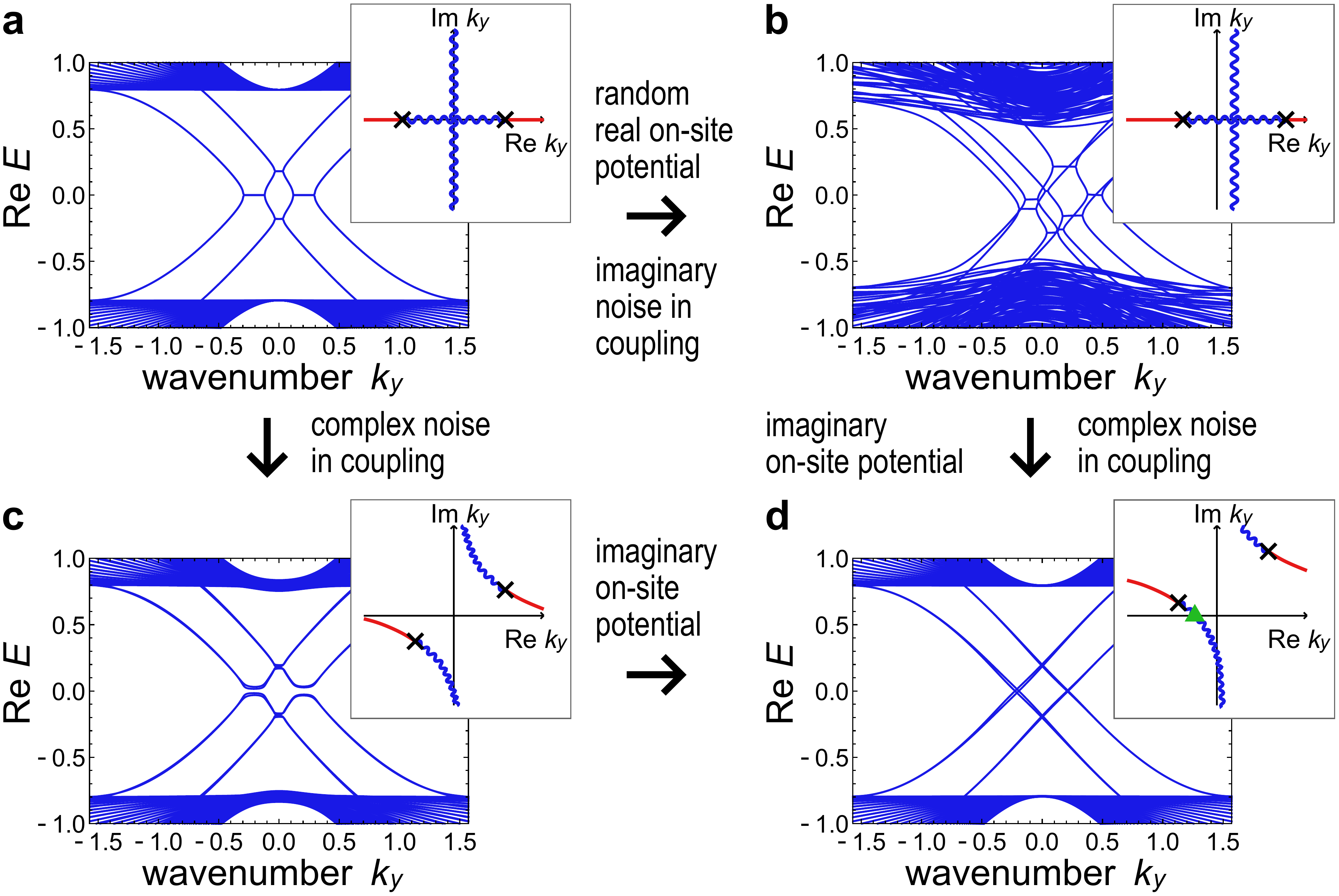}
\caption{\label{fig3}
Edge modes in disordered systems and their robustness. (a)-(d) Each inset shows the EPs (black crosses) and the curve of the degeneracy of the real (blue wave curves) and imaginary (red curves) parts of the edge eigenenergies in the complex wavenumber plane. Each inset corresponds to all the pairs of the edge modes with positive and negative group velocities. Because of the periodicity in the $y$ direction, the band structure corresponds to the behavior on the ${\rm Re}\,k_y$ axis in the insets, while the EPs and the degeneracy curves in the complex wavenumber space are useful for predicting the behavior of the edge modes. (a) The main panel shows the obtained exceptional edge modes in the system without disorder. The parameters used are $u=-1$, $c=0.2$, $\beta=0.14$, $\beta'=0.06$, and $\gamma=0.05$. (b) The main panel shows the edge dispersion with on-site random real potentials and imaginary disorder in the coupling term. There still exist EPs and edge modes in the bulk energy gap. The noise widths are set to be $W=0.5$ ($W=0.02$) for the random real on-site potential (the imaginary noise in the non-Hermitian coupling). (c) The gap is opened and the edge modes no longer exist in the system with random Hermitian couplings. The noise width is set to be $W=0.1$. (d) When we add imaginary on-site potentials, the edge modes are recovered even under the random Hermitian couplings. This is because the wavenumber should be real due to the periodic boundary condition in the $y$ direction and the real axis of the wavenumber plane crosses the degeneracy curve for the real parts of the eigenenergies (cf. the green triangle in the inset). However, EPs disappear from the edge modes. The noise width is the same as in panel (c) and the strength of the on-site imaginary potential is $g=0.2$.}
\end{figure*}

{\it Effective edge Hamiltonian and robustness of exceptional edge modes.---}
Robustness against the perturbation and the disorder is an important feature of topological edge modes. We note that conventional topological edge modes are fragile under the perturbations breaking the symmetry. To see what types of perturbations can sustain stable exceptional edge modes, we introduce a general one-dimensional effective Hamiltonian parametrized by wavenumber $k_y$,
\begin{equation}
 H_{\rm edge}(k_y) = \left(
  \begin{array}{cc}
   E_0 +k_y +\alpha & i\beta+\gamma \\
   i\beta -\gamma & E_0 -k_y-\alpha
  \end{array}
  \right) \label{general-edge},
\end{equation}
which describes the generic behavior of the low energy dispersion of edge modes. The diagonal elements represent the linear dispersion of two edge modes without couplings, and the off-diagonal parts represent the non-Hermitian coupling. The case of $\alpha=\gamma={\rm Im}\,\beta=0$ represents exceptional edge modes in the disorder-free system (${\rm Re}\,\beta\neq0$ is necessary to generate exceptional points in exceptional edge modes). The effective Hamiltonian has the eigenenergy $E^{\pm}(k_y) =E_0\pm \sqrt{(k_y+\alpha)^2-\beta^2-\gamma^2}$ and EPs at $k_y=-\alpha\pm \sqrt{\beta^2+\gamma^2}$. The topological index associated with EPs can guarantee their presence in the complex wavenumber space in this case (see Supplementary Information).

Since we consider the bulk gaps for the real parts of eigenenergies, the gapless edge modes remain when there exists a real wavenumber $k_y$ that satisfies ${\rm Re}\sqrt{(k_y+\alpha)^2-\beta^2-\gamma^2}=0$. Thus, we can conclude that $|{\rm Im}\sqrt{\beta^2+\gamma^2}| \leq |{\rm Im}\,\alpha|$ is a necessary and sufficient condition to realize robust edge modes (see Supplementary Information). Meanwhile, the exceptional edge modes remain when $-\alpha\pm \sqrt{\beta^2+\gamma^2}$ is real, i.e., (i) ${\rm Im}\,\alpha =0$, (ii) ${\rm Im}\,(\beta^2+\gamma^2) =0$, and (iii) ${\rm Re}\,(\beta^2+\gamma^2) >0$. From conditions (ii), (iii), we can derive ${\rm Im}\sqrt{\beta^2+\gamma^2}=0$ and thus can confirm that the condition for gapless edge modes must be satisfied under the condition for exceptional edge modes. We note that conditions (i), (ii) are equivalent to the condition for pseudo-Hermiticity \cite{Mostafazadeh2002}, which ensures that the eigenenergies are either real or pairs of complex conjugate values (i.e., $(k_y+\alpha)^2-\beta^2-\gamma^2$ is real in the present case). In general, disorder in the existing terms satisfies this condition. Meanwhile, nonzero ${\rm Re}\,\alpha$ breaks the time-reversal symmetry defined as $T=\sigma_y$, $TH(k_y)T^{-1}=H^{\ast}(-k_y)$. However, perturbation to ${\rm Re}\,\alpha$ does not affect the stability of exceptional edge modes even if they accompany nontrivial bulk topology of a time-reversal-symmetric system.

We can also relate the robustness of the exceptional edge modes to the symmetry and the topology of EPs. In one-dimensional systems, EPs can robustly exist under the $PT$ symmetry, the $CP$ symmetry, the pseudo-Hermiticity, or the chiral symmetry \cite{Okugawa2019,Budich2019,Kawabata2019b}. In the case of $E_0=\alpha=\gamma={\rm Im}\,\beta=0$, the effective edge Hamiltonian exhibits the $PT$ symmetry $PT=\sigma_z$, $PTH(k_y)(PT)^{-1}=H^{\ast}(k_y)$, the $CP$ symmetry $CP=\sigma_x$, $CPH(k_y)(CP)^{-1}=-H^{\ast}(k_y)$, the pseudo-Hermiticity $\eta=\sigma_z$, $\eta H(k_y)\eta^{-1}=H^{\dagger}(k_y)$, and the chiral symmetry $\Gamma=\sigma_x$, $\Gamma H(k_y)\Gamma^{-1}=-H^{\dagger}(k_y)$. We note that the pseudo-Hermiticity here is in a narrower class than that considered in the previous paragraph, that is, the operator $\eta$ is restricted to a local operator that only acts on the inner degrees of freedom. To preserve the $PT$ symmetry and/or the chiral symmetry, we need ${\rm Im}\,\alpha={\rm Im}\,\beta={\rm Re}\,\gamma=0$. We can also confirm that the preservation of the $CP$ symmetry and/or the pseudo-Hermiticity requires ${\rm Im}\,\alpha={\rm Im}\,\beta={\rm Im}\,\gamma=0$. From these equations, we can derive conditions (i), (ii) for realizing the exceptional edge modes discussed above and thus confirm that the exceptional edge modes robustly exist under sufficiently small $|{\rm Im}\,\gamma|$ and one of the following symmetry: the $PT$ symmetry, the $CP$ symmetry, the chiral symmetry, or the pseudo-Hermiticity. If we increase $|{\rm Im}\,\gamma|$, two exceptional points coalesce at critical strength of $|{\rm Im}\,\gamma|$, and the exceptional edge modes disappear under larger $|{\rm Im}\,\gamma|$. We can expect that the symmetry in the effective edge Hamiltonian is the same as that in the bulk and thus can utilize the symmetry as the guiding principle to predict what types of disorders remain exceptional edge modes.

To confirm the robustness of the exceptional edge modes in our model, we calculate the band structure with adding disorder (see Fig.~\ref{fig3}). We show that the exceptional edge modes still exist robustly under certain types of disorders, i.e., the random real on-site potential and the imaginary noise in the coupling terms (see Supplementary Information for details). These disorders preserve the modified $PT$ symmetry $P'T H(k_x,k_y) (P'T)^{-1} = H^{\ast}(-k_x,k_y)$ that plays the same role as the $PT$ symmetry in the edge band structure. Thus, the result is consistent with the discussion in the previous paragraph. Also, the on-site non-Hermitian term, $ig \sigma_z\otimes I\otimes I$, recovers the robustness of the edge modes against the real noise in the coupling terms, which lifts the degeneracy in the edge bands without on-site terms. With the on-site non-Hermitian term, since the two edge modes avoid each other in the imaginary part of the energy, they are not degenerate and thus are prohibited to open the real gaps (similar feature has been observed in the previous study \cite{Zhao2019} at the interface between gain and loss regions). This avoidance protects the edge modes from opening gaps as understood from the perturbation theory (see Supplementary Information for details). These results are consistent with the analysis of the effective edge Hamiltonian. In Supplementary Information, we further discuss the symmetry of the disordered Hamiltonian and clarify its relation to the robustness of the exceptional edge modes in both topologically trivial and nontrivial systems. Especially, we demonstrate the existence of the chiral-symmetry-protected exceptional edge modes and the importance of the modification of the $PT$ symmetry and the $CP$ symmetry for the protection of the exceptional edge modes.

\begin{figure*}
  \includegraphics[width=140mm,bb=0 0 890 610,clip]{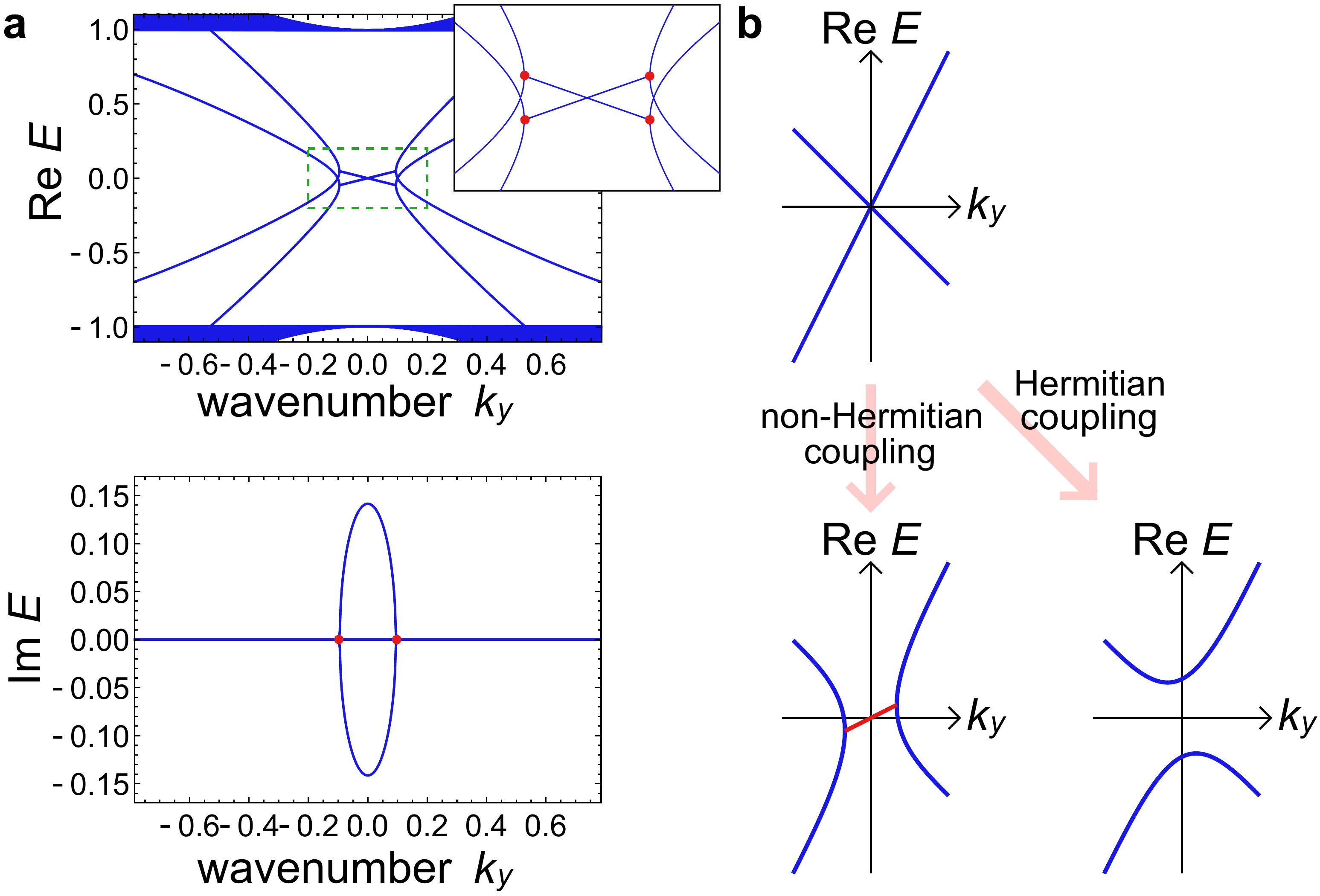}
\caption{\label{fig4}
Band structure for realizing lasing edge modes with nonzero group velocities. (a) Combining the Qi-Wu-Zhang (QWZ) model with large hoppings and the time-reversal QWZ model by a non-Hermitian coupling, we obtain the model with the edge band structure shown in the main panels. The edge dispersions between the pairs of EPs (red points) exhibit the nonzero imaginary part of the energy and the nonzero slope of the real part of the energy. Since the slope of the real part of the energy corresponds to the group velocity of the edge mode, these lasing edge modes have nonzero group velocity and propagate along the edge of the sample. The inset presents the enlarged view of the low momentum region indicated by the green dashed box and the red points represent the EPs in the edge dispersions. The parameters used are $u=-1$, $\beta=0.2$, and $\beta'=0.1$. (b) To obtain the lasing edge modes with nonzero group velocity, we utilize two edge modes which have opposite signs and different absolute values of the slope of the dispersion relation. By combining these edge modes by a non-Hermitian coupling term, we obtain exceptional edge modes, which can be applied to construct a topological insulator laser while Hermitian couplings open the gap due to the avoided crossing.}
\end{figure*}

\begin{figure*}
  \includegraphics[width=140mm,bb=10 5 880 290,clip]{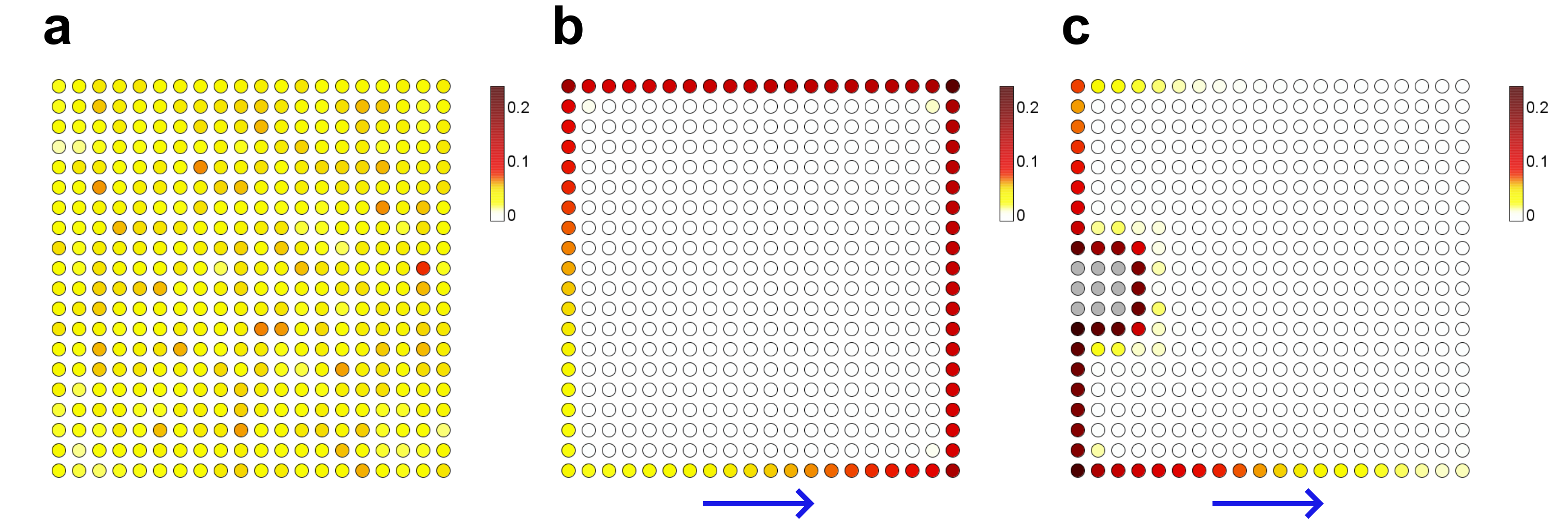}
\caption{\label{fig5}
Real-space simulation of the lasing edge modes. (a)-(c) The color represents the probability amplitude at each site, which is normalized so that the sum of squares gives unity. We set the parameter $u=-1$ throughout the calculations in this Figure. (a) Without the non-Hermiticity ($\beta=\beta'=0$), all the bulk modes can survive for a long time and thus lasing of edge modes fails to happen. (b) With the non-Hermitian coupling ($\beta=0.2$, $\beta'=0.1$), only the edge mode is enhanced even if we start from a random initial state. We can also confirm the propagation of the wave packet. (c) Starting from the excitation of one site at the edge, we confirm that a wave packet propagates along the edge of the sample. Even in the presence of distortion at the edge (represented by the grey sites), the edge wave packet avoids it and propagates without dissipation. The parameters used are $\beta=0.9$ and $\beta'=0.8$.}
\end{figure*}

In general, in Hermitian systems, the physical significance of the periodic table obtained from the bulk band topology is guaranteed by the bulk-edge correspondence that consistently predicts the presence or absence of robust gapless edge modes at open boundaries \cite{Hasan2010}. In contrast, in non-Hermitian cases, our findings force us to fundamentally alter this point of view. In particular, when $g\neq 0$ and $\gamma=0$ in our model, there exist the robust gapless edge modes as in Fig.~\ref{fig3}d ($g\neq0$ and $\gamma\neq0$), while the bulk topological invariant is trivial as inferred from the topological classification \cite{Gong2018,Zhou2019,Kawabata2019} (see Supplementary Information for details). In other words, the robust gapless edge modes found here violate the bulk-edge correspondence and cannot be captured by the existing periodic tables \cite{Gong2018,Zhou2019,Kawabata2019} of non-Hermitian topological phases, thus challenging the conventional classification based on Bloch Hamiltonians.

{\it Application to amplifying edge modes.---}
Next we show that amplified exceptional edge modes with nonzero group velocity can be realized. Specifically, we find that the general form of effective edge Hamiltonians is given by
\begin{equation}
 H_{\rm edge} = \left(
  \begin{array}{cc}
   E_0 -ia\partial_y & i\beta \\
   i\beta' & E_0 +i\partial_y
  \end{array}
  \right),\ \ a\neq 1,\ \beta\beta'>0 \label{1d-lasing}.
\end{equation}
We derive the dispersion relation of this effective Hamiltonian, $E(k_y)=E_0 + [(a-1)k_y \pm \sqrt{(a+1)^2 k_y^2 -4\beta\beta'} ]/ 2$, which exhibits exceptional points at $k_y=\pm2\sqrt{\beta\beta'}/(a+1)$ and nonzero group velocity. While this Hamiltonian describes the generic behavior of lasing edge modes utilizing exceptional edge modes, we construct a concrete tight-binding model represented by the following Hamiltonian: 
\begin{equation}
 H = \left(
  \begin{array}{cc}
   2H_{\rm QWZ} & i \beta \sigma_x \\
    i \beta' \sigma_x & H^{\ast}_{\rm QWZ}  
  \end{array}
  \right) \label{lasing-model},
\end{equation}
where $H_{\rm QWZ}$ is the Hamiltonian of the QWZ model. Figure \ref{fig4}a shows the edge band structure of this system. Nonzero imaginary parts of the eigenenergies appear only in the edge modes as in our first model. Also, the edge modes exhibit nonzero slopes of the real energy dispersion $\partial \mathrm{Re} E /\partial k$, which correspond to nonzero group velocities. Thus, we can observe the amplified wave packet propagating along the edge of the sample, which allows us to stably transfer the energy and thus may find potential applications. We note that this Hamiltonian is neither time-reversal symmetric nor pseudo-Hermitian. The sum of the Chern numbers for the bands under the energy gap is zero in our model, which indicates the bulk triviality in the conventional sense. Therefore, the edge modes are protected not by the bulk band topology but by the EPs.

\begin{figure}
  \includegraphics[width=84mm,bb=0 0 430 640,clip]{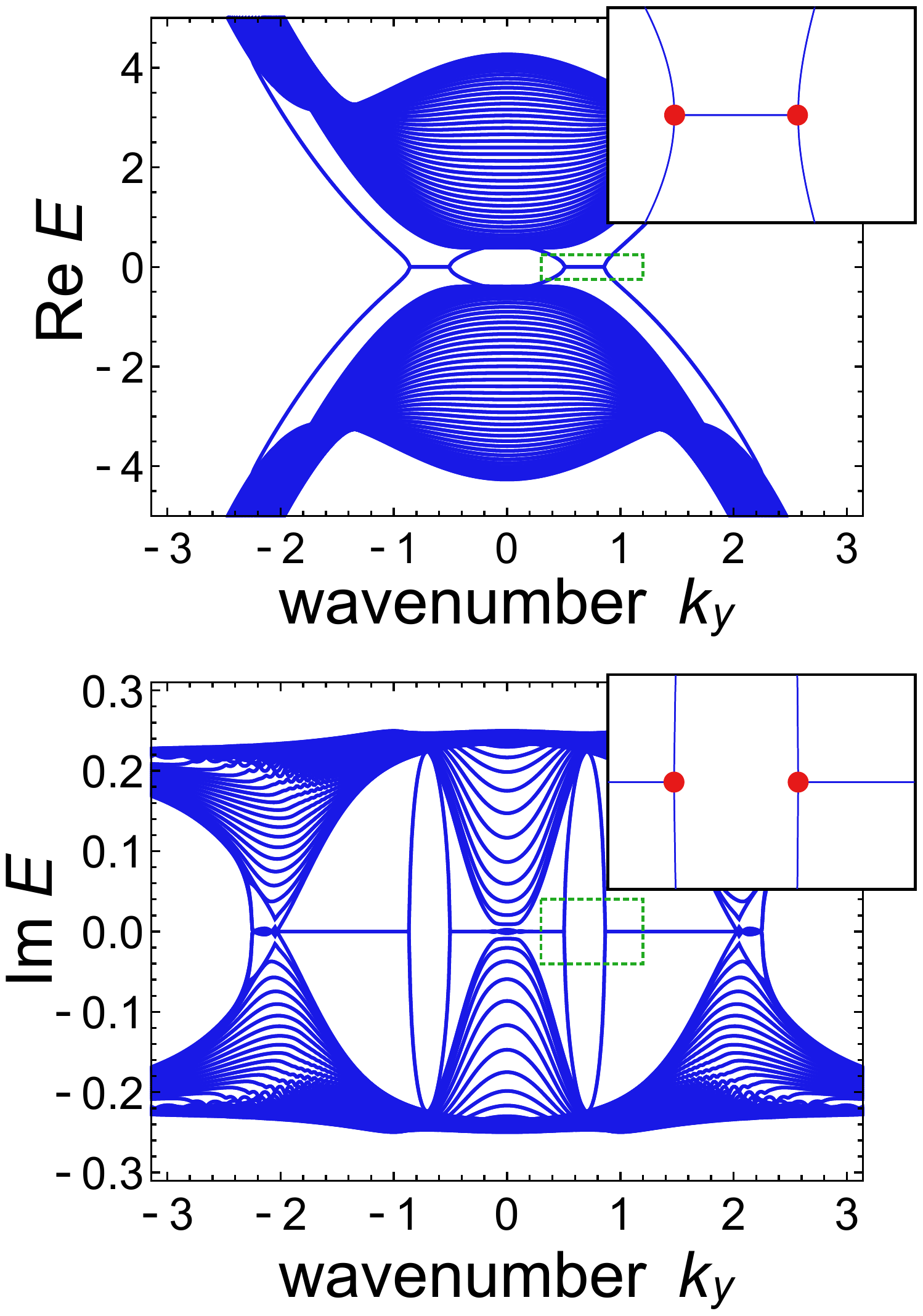}
\caption{\label{fig6}
Exceptional edge modes in the continuum model. The model is constructed from the continuum system with the Chern number $2$ and its time reversal combined by a non-Hermitian coupling. We discretize the space and the equation, and numerically calculate the band structure under the open boundary condition in the $x$ direction and the periodic boundary condition in the $y$ direction. While the model has topologically trivial bulk, it exhibits four edge bands per edge (doubly degenerated) protected by the exceptional points around ${\rm Re}\,E=0$ (indicated by the red points in the inset). The parameter used here are $M=0.5$, $\beta=0.5$, $a=1$, $b=0.3$, and $b'=0.2$.
}
\end{figure}

In general, exceptional edge modes are essential for this construction of a topological insulator laser. To obtain lasing edge modes, we must utilize a pair of edge modes localized at the same side whose dispersion relations cross each other without coupling terms. Also, to accomplish nonzero group velocity, the absolute values of the slopes of the edge energy bands must be different. Therefore, the degeneracy is not protected by the bulk band topology or the symmetry and thus can be resolved by Hermitian couplings as shown in Fig.~\ref{fig4}b. On the contrary, non-Hermitian couplings lead to both the enhancement and the robustness of the edge mode. Thus, the lasing edge mode must be an exceptional edge mode.

We demonstrate the real-space dynamics of our topological insulator laser (Eq.~\eqref{lasing-model}) by numerical calculations (see Supplementary Videos 1-3). Figure \ref{fig5} shows the snapshots for the real-space distributions of the probability densities of the wave functions. Without non-Hermitian coupling, $\beta = \beta' = 0$, the bulk oscillation survives. With non-Hermitian coupling, $\beta,\beta' \neq 0$, the bulk oscillation becomes much smaller than the edge oscillation in a short time, and only the edge mode remains even if we start with the random initial state. Also, we can confirm that the edge mode has nonzero group velocity. Furthermore, we introduce disorder on the edge and excite only one edge site. Then, we obtain the propagating edge mode without backscattering. This implies the robustness of the exceptional edge mode against the disorder at the edge. In contrast to the previous research \cite{Harari2018,Bandres2018}, we do not need to introduce judicious gain along the edge. This difference can potentially facilitate the realization of topological insulator laser in various physical setups.

\begin{figure*}
  \includegraphics[width=140mm,bb=0 0 450 430,clip]{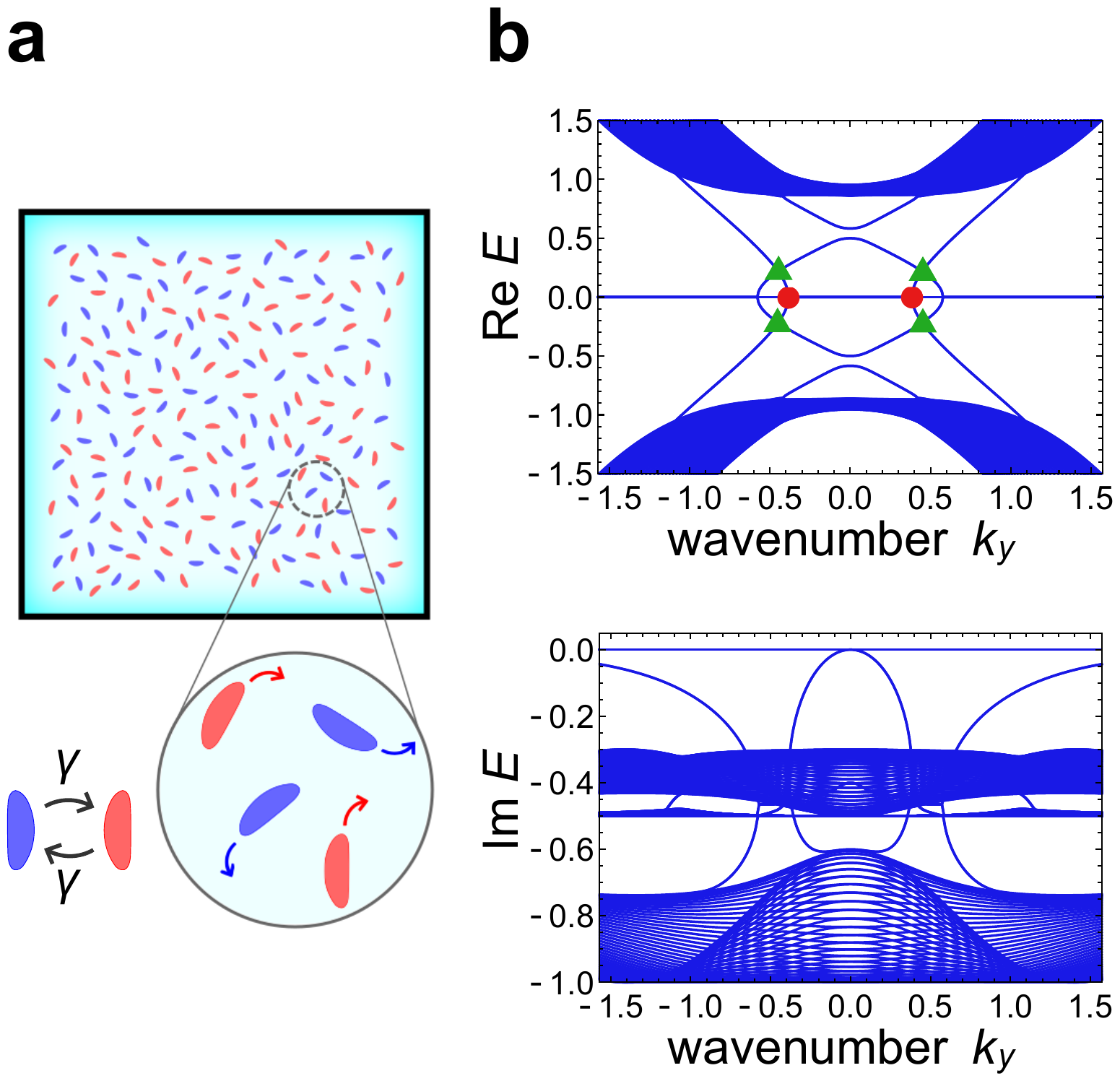}
\caption{\label{fig7}
Exceptional edge modes in chiral active matter. (a) Two-component chiral active matter is considered, where each active particle tends to move in the left (right) direction as shown in the blue (red) arrow. These two types of chiral active matter are mixed and the chirality can flip at a constant rate. (b) The edge band structure calculated for the cylindrical system is shown. The negativity of the imaginary part of eigenenergies implies the linear stability of the steady state. Since there exist four bands per edge in the bulk gap, the bulk topology is trivial in the conventional sense. The red circles represent the EPs that lead to the robustness of the edge modes. Also, we can find apparent edge-band crossings in the bulk real energy gap, which are denoted as the green triangles. However, the edge modes do not really cross there, because they avoid each other in their imaginary parts of the eigenenergies. Therefore, the apparent degeneracies are artifacts, and thus the gapless edge modes are not broken by disorder. The other crossings on the ${\rm Re}\,E=0$ axis correspond to the points where the edge bands come out of the bulk bands. The parameters used are $\omega_0=1$, $\nu^o=0.5$, $\gamma=0.3$, and $\beta=0.5$.
}
\end{figure*}

{\it Active matter realization of exceptional edge modes.---}
Analogous to the conventional topological edge modes \cite{Xu2011}, the exceptional edge modes can also exist in continuum systems. We construct a continuum toy model and confirm the existence of exceptional edge modes by calculating the band structure (see Supplementary Information for the detail of the model). Figure \ref{fig6} represents the edge band structure of the continuum model. While the bulk bands are topologically trivial as in the tight-binding model (Eq.~\eqref{toy-model1}), it exhibits the robust exceptional edge modes.

To show that exceptional edge modes are indeed realizable in realistic systems, we focus on a continuum active matter model. We consider chiral active matter without the left-right symmetry in which each particle moves on a clockwise (or counterclockwise) circular trajectory (Fig.~\ref{fig7}a). We mix clockwise and counterclockwise moving particles. We also assume that the chirality of active particles flips occasionally and the flipping rate $\gamma$ is symmetric between clockwise and counterclockwise moving particles. The active particles have long and narrow shapes. We assume that polar interaction acts on them, which aligns the neighboring particles and effectively appears in some self-propelled rods \cite{Bar2020}. Anti-polar interaction is also allowed to exist between particles with opposite chirality. This setup can possibly be experimentally realized by utilizing bacteria \cite{DiLuzio2005}, artificial L-shaped particles \cite{Kummel2013}, or robotic rotors \cite{Yang2019}. To be concrete, we expect that exceptional edge modes can appear in bacteria swimming between the two plates at the distance shorter than the bacteria length and in L-shaped active particles that are occasionally turned over (see Supplementary Information for further details). Here, the crucial requirements for the experimental realization of exceptional edge modes are the flippable chirality and the momentum coupling.

In Fig.~\ref{fig7}b, we show the existence of exceptional edge modes by numerically diagonalizing the effective Hamiltonian of our active matter model, which is derived by linearizing the hydrodynamic equations \cite{Marchetti2013,Toner1995} (see Supplementary Information). We confirm that a pair of EPs appear at the frequency $\omega=0$ and support the robustness of the edge modes. Meanwhile, at a glance, there are degeneracies in the bulk gap. However, the edge modes avoid each other in the imaginary part of the frequency like in Fig.~\ref{fig3}d, and thus these apparent degeneracies are robust against the disorder. We obtain two other crossings on the ${\rm Re}\,E=0$ axis, which correspond to the points where unprotected edge bands appear from the bulk bands around the axis (see Supplementary Information for the detail on the function of these crossing). In realistic experimental situations, we expect that the oscillation of the fluctuation of the density or the velocity field propagates at the edge of the sample in the direction depending on the chirality of particles (i.e., clockwise or counterclockwise) when we apply the perturbation with a small frequency compared to the bulk bandgap, which is almost equal to the frequency of rotation $\omega$. The imaginary parts of the eigenvalues are all nonpositive, and thus we need further modification of the active system to apply the proposed setup to lasing devices.

{\it Summary and Discussions.---}
We revealed the existence of robust gapless edge modes unique to non-Hermitian systems by utilizing EPs. These edge modes, which we called exceptional edge modes, can exist even when the bulk topology is trivial. We also analyzed and confirmed the robustness of the edge modes by constructing the effective edge Hamiltonian. By utilizing these edge modes, we proposed a topological insulator laser whose edge modes were amplified and propagate along the edge. We also showed that the chiral active particles with chirality flipping can exhibit the exceptional edge modes and thus they can be realized in the upcoming experimental techniques of active matter, while the model analyzed here has only nonpositive imaginary parts of the eigenfrequencies and thus exhibits no lasing behavior.

The edge modes found here provide an alternative design principle to realize scattering-free edge current intrinsic to non-Hermitian systems, which is not based on the bulk topology, and thus indicates that the conventional arguments on bulk topology, including the periodic tables \cite{Gong2018,Zhou2019,Kawabata2019}, are insufficient to predict the presence or absence of robust edge modes in non-Hermitian systems. Exceptional edge modes in higher-dimensional systems are important to further elucidate a nontrivial role of open boundaries in non-Hermitian systems. Furthermore, our active matter model demonstrates that hydrodynamics of active matter can be applied to non-Hermitian topological phenomena, indicating that active matter provides a useful platform for exploring non-Hermitian topology.

We thank Zongping Gong, Kohei Kawabata, Kyogo Kawaguchi, Daiki Nishiguchi, Shun Otsubo, Kazumasa Takeuchi, and Hiroki Yamaguchi for valuable discussions. K.S. is supported by World-leading Innovative Graduate Study Program for Materials Research, Industry, and Technology (MERIT-WINGS) of the University of Tokyo. Y.A. is supported by JSPS KAKENHI Grant Numbers JP16J03613 and JP19K23424. T.S. is supported by JSPS KAKENHI Grant Numbers JP16H02211 and JP19H05796.

\widetext
\pagebreak
\begin{center}
\textbf{\large Supplementary Materials}
\end{center}

\renewcommand{\theequation}{S\arabic{equation}}
\renewcommand{\thefigure}{S\arabic{figure}}
\setcounter{equation}{0}
\setcounter{figure}{0}

\subsection{Derivation of the condition for remaining gapless edge modes}
From the analysis of the effective edge Hamiltonian (cf. Eq.~(3) in the main text), we conclude that $|\rm{Im}\sqrt{\beta^2+\gamma^2}| \leq |\rm{Im}\,\alpha|$ is the necessary and sufficient condition for the existence of gapless edge modes. To show this, we start from the equation ${\rm Re}\sqrt{(k_y+\alpha)^2-\beta^2-\gamma^2}=0$, which leads to the eigenenergy $E^{\pm} = E_0 \pm i \delta$ with $\delta$ being a real number. For a wavenumber satisfying this equation, the real parts of the two eigenenergies become the same, and thus we obtain the gapless edge modes. Then, we prove that the existence of such a wavenumber is equivalent to the condition $|{\rm Im}\sqrt{\beta^2+\gamma^2}| \leq |{\rm Im}\,\alpha|$. We consider $\alpha'={\rm Im}\,\alpha$ and $k_y' = k_y + {\rm Re}\,\alpha$, and obtain $(k_y+\alpha)^2 = k_y'^2-\alpha'^2+2ik_y'\alpha'$. Similarly, if we describe $\sqrt{\beta^2+\gamma^2}=l+i\beta'$, we obtain $\beta^2+\gamma^2 = l^2-\beta'^2+2il\beta'$. To make $\sqrt{(k_y+\alpha)^2-\beta^2-\gamma^2}$ a real or a pure imaginary number, we have to set ${\rm Im}[ (k_y+\alpha)^2-\beta^2-\gamma^2] = 0$ and thus consider $k_y' = l\beta'/\alpha'$. Then, we obtain
\begin{equation}
{\rm Re}[ (k_y+\alpha)^2-\beta^2-\gamma^2] = (l^2+\alpha'^2)(\beta'^2-\alpha'^2)/\alpha'^2,
\end{equation} 
and this is zero or negative if and only if $\beta'^2 \leq \alpha'^2$. Thus $|{\rm Im}\sqrt{\beta^2+\gamma^2}| \leq |{\rm Im}\,\alpha|$ is the necessary and sufficient condition for the existence of a wavenumber satisfying ${\rm Re}\sqrt{(k_y+\alpha)^2-\beta^2-\gamma^2}=0$.

\subsection{Tight-binding models in real-space basis}
Here we explicitly describe the Hamiltonian used in the main text. The Hamiltonian of the Qi-Wu-Zhang (QWZ) model \cite{Qi2006} in the real-space basis is denoted as
\begin{eqnarray}
 H_{\rm QWZ} &=& \sum_x \sum_y \left( |x+1,y\rangle \langle x,y| \otimes \frac{\sigma_z + i \sigma_x}{2} + {\rm h.c.} \right) \nonumber\\
 &{}& + \sum_x \sum_y \left( |x,y+1\rangle \langle x,y| \otimes \frac{\sigma_z + i \sigma_y}{2} + {\rm h.c.} \right)  + u \sum_x \sum_y |x,y\rangle \langle x,y| \otimes \sigma_z, \label{QWZ}
\end{eqnarray}
where ${\rm h.c.}$ represents the Hermitian conjugate of the previous term and $\sigma_i$ is the $i$th component of the Pauli matrices. This model contains two sublattices. To construct a minimal model for exceptional edge modes, we prepare two layers of the QWZ model to construct a higher Chern number insulator as is done in the previous studies \cite{Trescher2012,Liu2012}. Then, we combine them with their time-reversal counterparts by non-Hermitian and Hermitian couplings. The obtained Hamiltonian can be written by a $4\times4$ matrix form as
\begin{equation}
 H = \left(
  \begin{array}{cccc}
   H_{\rm QWZ} & cI_2 & i a\beta \sigma_x & -i\gamma\sigma_x \\
   cI_2 & H_{\rm QWZ} & i\gamma\sigma_x & i a\beta' \sigma_x \\
    i a\beta \sigma_x & -i\gamma\sigma_x & H^{\ast}_{\rm QWZ} & cI_2 \\
   i\gamma\sigma_x & i a\beta' \sigma_x & cI_2 & H^{\ast}_{\rm QWZ} 
  \end{array}
  \right). \label{toy-model2}
\end{equation}
In the numerical calculations of Fig.~1b and 3 in the main text, we use the parameters $u=-1$, $c=0.2$, $\beta=0.14$, $\beta'=0.06$, $\gamma=0.05$, and $a=1$. Also in the numerical calculation of Fig.~2 in the main text, we use the parameters $u=-1$, $c=0.2$, $\beta=0.9$, $\beta'=0.7$, and $\gamma=0.8$ and change the parameter $a$.

We also introduce the disorder terms in Fig.~3. We use $a(x)\{I_2,\sigma_z\}\otimes\{I_2,\sigma_z\}\otimes\{I_2,\sigma_z\}$ and $b(x)\{\sigma_x,i\sigma_y\}\otimes\{I_2,\sigma_z\}\otimes\{\sigma_x,i\sigma_y\}$ as the random real on-site potential and the imaginary and real noise in the non-Hermitian coupling for each, where brackets mean that we introduce all the combinations made by choosing either one in each bracket. $a(x)$ and $b(x)$ are random values for each $x$ from uniform distributions ranging $[-W,W]\in\mathbb{R}$ or $i[-W,W]\in i\mathbb{R}$. We set $a(x)$ to be real and $W=0.5$ in Fig.~3b. Also, we set $b(x)$ to be imaginary (real) for the imaginary (real) noise in the non-Hermitian coupling and $W=0.02$ ($W=0.1$) in Fig.~3b (Fig.~3c,d). We also consider on-site imaginary potential $ig\sigma_z\otimes I_2\otimes I_2$ and set $g=0.2$ in Fig.~3d.

\subsection{Absence of imaginary parts of eigenenergies in the two-layered Bernevig-Hughes-Zhang model}
\begin{figure}[t]
\includegraphics[width=80mm,bb=0 0 480 270,clip]{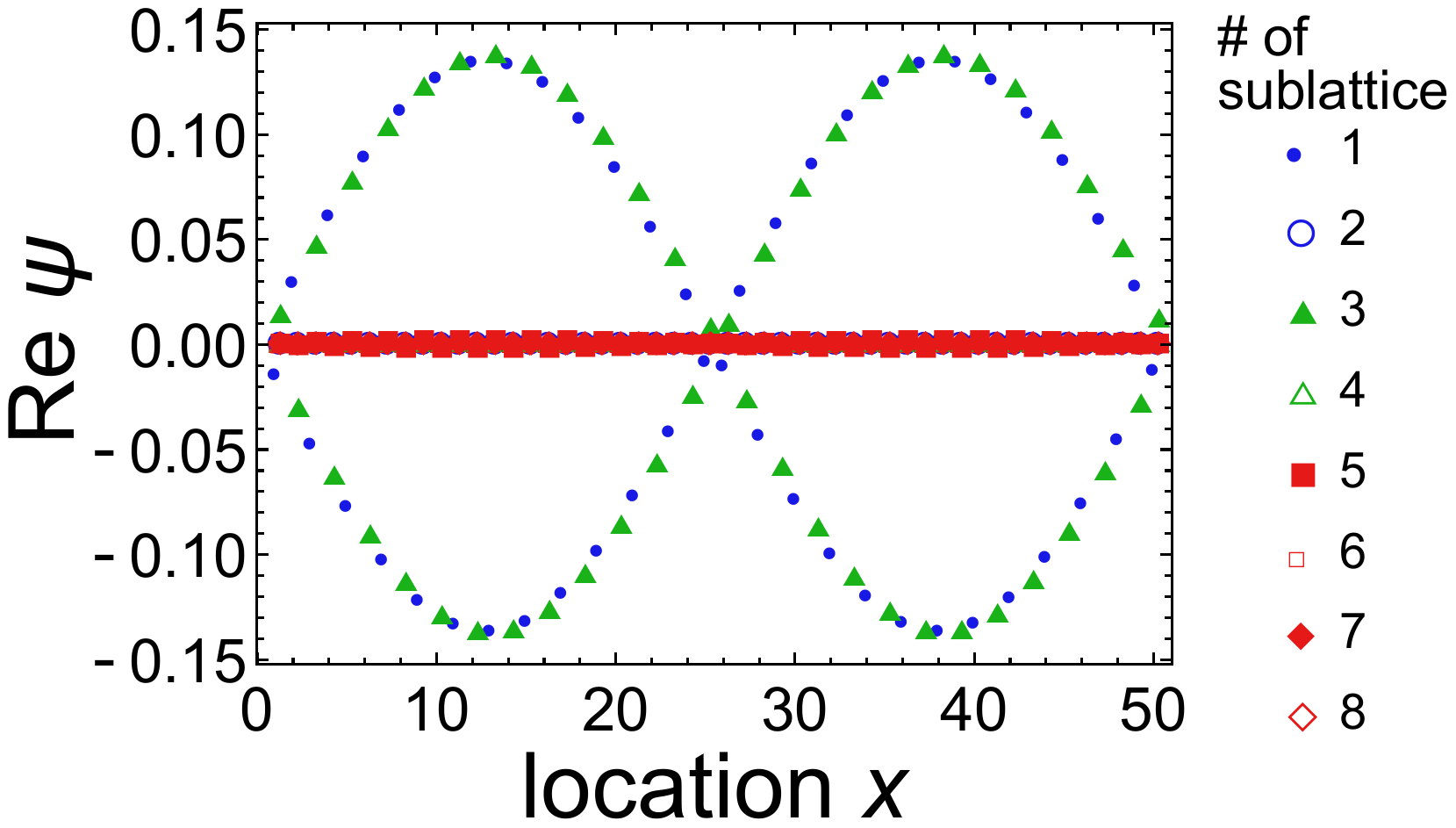}
\caption{\label{supplefig1}
Bulk eigenmode in the two-layered non-Hermitian Bernevig-Hughes-Zhang model. The bulk eigenmode is calculated under the open boundary condition in the $x$ direction and the periodic boundary condition in the $y$ direction. We align 50 sites in the $x$ direction. The legend shows the correspondence between each mark and the sublattice (eight per site). The parameters used are the same as in Fig. 1 in the main text.
}
\end{figure}
\noindent
We discuss why the bulk bands of the two-layered Bernevig-Hughes-Zhang model exhibit only the real eigenenergies (cf. Eq.~(1) and Fig.~1b in the main text). We can directly show that the imaginary parts of the bulk eigenenergies become zero at least in the range of the first-order perturbation with respect to the non-Hermitian coupling. If we describe the QWZ Hamiltonian as $H=\mathbf{R} \cdot \boldsymbol{\sigma}$, where $\mathbf{R} = (\sin k_x, \sin k_y, u+\cos k_x + \cos k_y)^T=|R| (\sin\theta \cos\phi, \sin\theta \sin\phi, \cos\theta)^T$ and $\boldsymbol{\sigma}= (\sigma_x,\sigma_y,\sigma_z)^T$, we obtain the explicit forms of the eigenvectors of the QWZ model,
\begin{equation}
 |+,\mathbf{R} (\mathbf{k}) \rangle = e^{-\frac{i\psi}{2}} \left(
  \begin{array}{c}
   e^{-\frac{i\phi}{2}} \cos \frac{\theta}{2} \\
   e^{\frac{i\phi}{2}} \sin \frac{\theta}{2}
  \end{array}
  \right),
\end{equation}
\begin{equation}
 |-,\mathbf{R} (\mathbf{k}) \rangle = e^{-\frac{i\psi}{2}} \left(
  \begin{array}{c}
   e^{-\frac{i\phi}{2}} \sin \frac{\theta}{2} \\
   - e^{\frac{i\phi}{2}} \cos \frac{\theta}{2}
  \end{array}
  \right).
\end{equation}
By utilizing these expressions, we can obtain the eigenvectors of the two-layered QWZ model,
\begin{equation}
 |+,\pm,\mathbf{R} (\mathbf{k}) \rangle = \frac{1}{\sqrt{2}} e^{-\frac{i\psi}{2}} \left(
  \begin{array}{c}
   e^{-\frac{i\phi}{2}} \cos \frac{\theta}{2} \\
   e^{\frac{i\phi}{2}} \sin \frac{\theta}{2} \\
   \mp e^{-\frac{i\phi}{2}} \cos \frac{\theta}{2} \\
   \mp e^{\frac{i\phi}{2}} \sin \frac{\theta}{2}
  \end{array}
  \right),
\end{equation}
\begin{equation}
 |-,\pm,\mathbf{R}(\mathbf{k}) \rangle = \frac{1}{\sqrt{2}} e^{-\frac{i\psi}{2}} \left(
  \begin{array}{c}
   e^{-\frac{i\phi}{2}} \sin \frac{\theta}{2} \\
   - e^{\frac{i\phi}{2}} \cos \frac{\theta}{2} \\
   \mp e^{-\frac{i\phi}{2}} \sin \frac{\theta}{2} \\
   \pm e^{\frac{i\phi}{2}} \cos \frac{\theta}{2}
  \end{array}
  \right).
\end{equation}
On the other hand, we note $H^{\ast}_{\rm QWZ} (-\mathbf{k}) = \mathbf{R}'(\mathbf{k}) \cdot \boldsymbol{\sigma}$ with $\mathbf{R}' (\mathbf{k}) = (-\sin k_x, \sin k_y, u+\cos k_x+\cos k_y)$. If we rewrite $\mathbf{R}'$ as $|R'| (\sin\theta' \cos\phi', \sin\theta' \sin\phi', \cos\theta')$, we obtain $\theta' (\mathbf{k}) = \theta (\mathbf{k})$, $\phi' (\mathbf{k}) = \pi - \phi (\mathbf{k})$. The first-order perturbation by non-Hermitian and Hermitian couplings can be obtained from a matrix whose elements are $\langle \pm ,\mathbf{R}(\mathbf{k}) | \sigma_x | \pm, \mathbf{R}' (\mathbf{k}) \rangle$, and we can calculate them as, for example,
\begin{eqnarray}
 \langle + ,\mathbf{R}(\mathbf{k}) | \sigma_x | +, \mathbf{R}' (\mathbf{k}) \rangle &=& e^{i\frac{\phi+\phi'}{2}} \cos\frac{\theta}{2} \sin\frac{\theta'}{2} + e^{-i\frac{\phi+\phi'}{2}} \sin\frac{\theta}{2} \cos\frac{\theta'}{2} \nonumber\\
 &=& (e^{i\frac{\pi}{2}} + e^{-i\frac{\pi}{2}}) \cos\frac{\theta}{2} \sin\frac{\theta}{2} = 0.
\end{eqnarray}
Therefore, the first-order perturbation becomes zero for the bulk bands and thus does not generate nonzero imaginary parts of the eigenenergies.

We can justify the use of the Bloch Hamiltonian for the purpose of predicting the bulk eigenstates under the open boundary condition, while recent studies \cite{Kunst2018,Xiong2018,Yao2018,Lee2019,Yokomizo2019} have revealed that some non-Hermitian systems breaking the bulk-edge correspondence exhibit bulk modes different from ones predicted from their Bloch Hamiltonians. We numerically calculate the bulk eigenstates and confirm the consistency with the prediction from the Bloch Hamiltonian. Supplementary Figure \ref{supplefig1} shows an example of bulk eigenstates. It resembles the Bloch wave (i.e. sine curve). Besides, its probabilistic density is localized at the half of sublattices, which implies that two bulk eigenmodes obtained from the QWZ model and its time-reversal counterpart are separated under the small non-Hermitian coupling. These behaviors of the bulk eigenstates are consistent with the discussion in the previous paragraph. Therefore, we can expect that the Bloch Hamiltonian predicts the behavior of the bulk eigenmodes in the model under the open-boundary condition. We note that the breakdown of the bulk-edge correspondence discussed here indicates the existence of the unpredicted robust edge modes with the exceptional points (EPs) and does not imply the other context of the breakdown such as non-Hermitian skin effect \cite{Yao2018,Lee2019,Yokomizo2019}. In the previous paragraph, furthermore, we use the bulk eigenstates predicted from the Bloch Hamiltonian as the eigenstates of the nonperturbed Hermitian Hamiltonian and thus can expect that the perturbation calculation based on such nonperturbed eigenstates reach the proper conclusion.

We can also partially explain the reality of the eigenenergies of the bulk modes from the $PT$ symmetry of the system. The $PT$ symmetry of the Hamiltonian $H(\mathbf{k})$ of the two-layered non-Hermitian Bernevig-Hughes-Zhang model is described as 
\begin{equation}
 PT = \sigma_x \otimes I_2 \otimes \sigma_z,\ \ PTH(\mathbf{k}) (PT)^{-1} = H^{\ast}(\mathbf{k}).
\end{equation}
The $PT$ symmetry guarantees that the eigenvalues become real or appear as pairs of complex conjugates. However, it is impossible to predict which of those behaviors appears.

\subsection{Exceptional edge modes in a nontrivial tight-binding model}
\begin{figure}[b]
\includegraphics[width=120mm,bb=0 0 1095 730,clip]{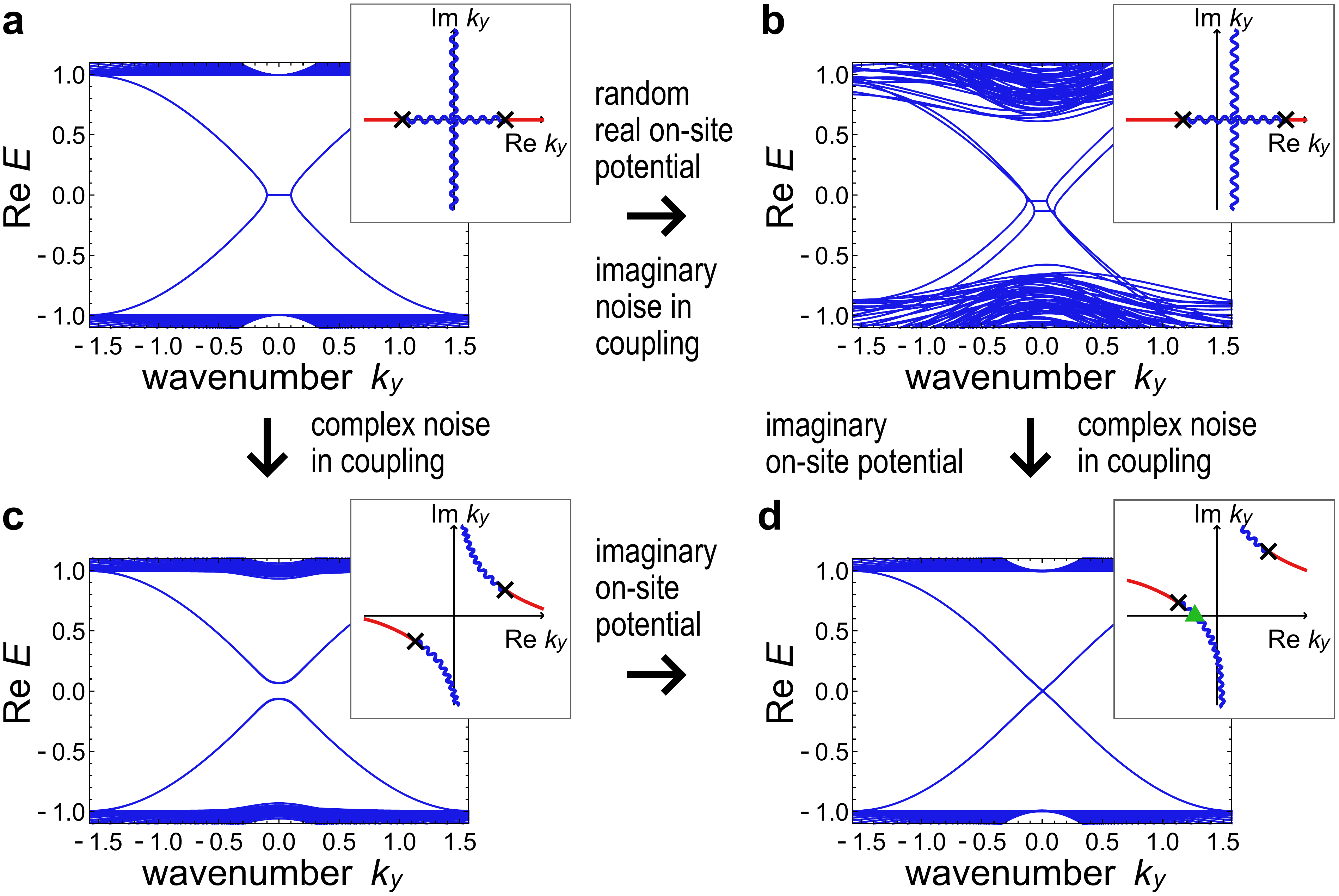}
\caption{\label{supplefig2}
Edge modes in disordered nontrivial systems and their robustness. (a)-(d) Each inset shows the exceptional points (EPs) (black crosses) and the curve of the degeneracy of the real (blue wave curves) and imaginary (red curves) parts of the edge eigenenergies in the complex wavenumber plane as in Fig.~3 in the main text. (a) The main panel shows the obtained exceptional edge modes in the nontrivial system (Supplementary Eq.~\eqref{nontrivial-model}) without disorder. The parameters used are $u=-1$ and $\beta=0.1$. (b) The main panel shows the edge dispersion with on-site random real potentials and imaginary disorder in the coupling term. There still exist EPs and edge modes in the bulk energy gap as in the model in the main text. The noise width is set to be $W=0.5$ ($W=0.02$) for the random real on-site potential (the imaginary noise in the non-Hermitian coupling). (c) The gap is opened and the edge modes no longer exist in the system with random Hermitian couplings. The noise width is set to be $W=0.1$ for the real noise in the non-Hermitian coupling. (d) When we add imaginary on-site potentials, the edge modes are recovered even under the random Hermitian couplings, while EPs disappear from the edge modes. The noise width for the real noise in the non-Hermitian coupling is the same as in panel (c) and the strength of the on-site imaginary potential is $g=0.2$.
}
\end{figure}
\noindent
Here, we discuss the existence and the robustness of exceptional edge modes in topologically nontrivial systems, while we concentrate on the systems with topologically trivial bulks in the main text. To construct the topologically nontrivial model, we again utilize the construction procedure for time-reversal-symmetric topological insulators. We combine the QWZ model and its time-reversal counterpart by a non-Hermitian coupling and obtain the Hamiltonian,
\begin{eqnarray}
 H &=& \left(
  \begin{array}{cccc}
   H_{\rm QWZ} & i \beta \sigma_x \\
   i \beta \sigma_x & H^{\ast}_{\rm QWZ} 
  \end{array}
  \right)\nonumber\\
  &=&  (u+\cos k_x + \cos k_y) I_2 \otimes \sigma_z + \sin k_y I_2 \otimes \sigma_y + \sin k_x \sigma_z \otimes \sigma_x + i\beta \sigma_x \otimes \sigma_x \label{nontrivial-model}
\end{eqnarray}
where $\beta$ is real. This Hamiltonian exhibits the time-reversal symmetry and the pseudo-Hermiticity, and we can obtain both the nontrivial $\mathbb{Z}_2$ and $\mathbb{Z}$ invariant defined in the time-reversal symmetric and the pseudo-Hermitian systems. ($\mathbb{Z}$ invariant can be calculated from the Chern number of $\eta H$ \cite{Gong2018,Kawabata2019}, where $\eta$ is the operator determining the pseudo-Hermiticity of the Hamiltonian $H$.) Therefore, the edge modes appear in this model are protected by the topology and the symmetries of the system. However, the EPs in the edge bands protect the edge modes from the gap openings against disorders breaking the time-reversal symmetry and the pseudo-Hermiticity. To confirm such protection, we numerically calculate the edge band structures under the existence of disorders. As in Fig.~3 in the main text, we consider a random real on-site potential, imaginary and real noises in the non-Hermitian coupling and an imaginary on-site potential expressed as follows:
\begin{eqnarray}
 &{}& a(x)\{I_2,\sigma_z\}\otimes\{I_2,\sigma_z\} \ \ \ \text{(random real on-site potential)} \\
 &{}& b(x)\{\sigma_x,i\sigma_y\}\otimes\{\sigma_x,i\sigma_y\} \ \ \ \text{(noise in the non-Hermitian coupling)} \\
 &{}& ig \sigma_z \otimes I_2 \ \ \ \text{(imaginary on-site potential)} 
\end{eqnarray}
where brackets mean that we introduce all the combinations made by choosing either one in each bracket. $a(x)$ and $b(x)$ are random values for each $x$ from uniform distributions ranging $[-W,W]\in\mathbb{R}$ or $i[-W,W]\in i\mathbb{R}$. We set $a(x)$ to be real and $b(x)$ to be imaginary (real) for the imaginary (real) noise in the non-Hermitian coupling. Supplementary Figure \ref{supplefig2} shows the results of the numerical calculations. We can confirm that all the results are consistent with the prediction from the effective edge Hamiltonian and the results of the numerical calculations on the disordered two-layered non-Hermitian Bernevig-Hughes-Zhang model (cf. Eq.~(3) and Fig.~3 in the main text). We note that the random real on-site potential and the imaginary noise in the non-Hermitian coupling breaks both the time-reversal symmetry and the pseudo-Hermiticity, while the exceptional edge modes exist robustly against these disorders. Therefore, exceptional edge modes are protected against symmetry-breaking disorder and thus improve the robustness of edge modes even in topologically nontrivial systems.

\subsection{Topological index for exceptional points in the complex wavenumber space}
It has been proposed \cite{Shen2018} that the topological invariants for EPs can be defined in the two-dimensional wavenumber space. In our models, however, EPs appear in the edge band structure for one-dimensional wavenumber $k_y$. Considering the complex wavenumber, we assume that the EPs appear in the two-dimensional parameter space of the real and imaginary parts of the wavenumber. Then, we can define the topological invariant for the EPs in the edge band structure as
\begin{equation}
 \nu_{\pm} = \frac{1}{2\pi} \oint_{|k_y \mp \beta|=\delta} \frac{d}{dk_y} \arg [E^{+}(k_y)-E^{-}(k_y)] dk_y \label{EPindex},
\end{equation}
where $E^{\pm} (k_y)$ is the eigenenergy of the two bands around the EPs and $\delta$ is the small radius of the circular integral path. We note that in our model two edge eigenstates coalesce at each EP, while in general, three or more edge eigenstates can coalesce at one EP. In the latter case, we need to generalize the definition of the topological index to include the effect of all the relevant eigenvalues.

We next calculate the topological index of the EPs in the exceptional edge modes via the effective edge Hamiltonian $H(k_y) = E_0 I + k_y \sigma_z + i\beta' \sigma_x$, which describes exceptional edge modes in the disorder-free system. The Hamiltonian possesses the eigenenergies $E^{\pm}(k_y)=E_0 \pm \sqrt{k_y^2-\beta'^2}=E_0 \pm \sqrt{k_y-\beta'}\sqrt{k_y+\beta'}$ and EPs at $k_y=\pm\beta'$. We can calculate the topological index around the EP at $k_y = \beta'$ as
\begin{eqnarray}
 \nu &=& \frac{1}{2\pi} \oint \frac{d}{dk_y} \arg [2\sqrt{k_y-\beta'}\sqrt{k_y+\beta'}] dk_y \nonumber\\
 &=& \frac{1}{2\pi} \int_0^{2\pi} \frac{d}{d\theta} \arg [2\sqrt{\delta e^{i\theta}}\sqrt{2\beta'+\delta e^{i\theta}}] d\theta \nonumber\\
 &=& \frac{1}{2\pi} \int_0^{2\pi} \frac{d}{d\theta} [\theta/2+\arg(2\beta'+\delta e^{i\theta})/2] d\theta \nonumber \\
 &=& \frac{1}{2},
\end{eqnarray}
where we consider a circle with a small radius $\delta$ centered at $k_y=\beta'$ as the integral path. Similarly, we can check that the topological index of the EP at $k_y = -\beta'$ is $1/2$. These topological indices are invariant under the continuum deformation of the Hamiltonian, thus indicating the robustness of the EPs in the complex wavenumber space.

\subsection{Perturbation analysis on the edge modes avoiding in the imaginary part of the energy}
Here, we discuss the origin of the robustness of the edge modes in Fig.~3d in the main text, which apparently cross each other but have the different imaginary parts of the eigenenergies. We utilize the perturbation theory and the continuity of the band structure. Since each edge band is continuous with respect to the wavenumber $k_y$ and the change of the parameter, edge bands connecting positive and negative bulk bands should remain gapless until two or more edge bands become degenerate. From the perturbation calculation, however, we can understand that such degeneracy is prohibited under a small disorder. Therefore, the gapless edge bands in Fig.~3d in the main text are robust against disorder.

To be concrete, we consider two edge modes, $|\psi_1(k_y)\rangle$ and $|\psi_2(k_y)\rangle$ with the eigenvalues $E_1(k_y)$ and $E_2(k_y)$, for each. We assume that the other eigenvalues are separated far from $E_1(k_y)$ and $E_2(k_y)$ and thus the other eigenvectors only have a negligible effect on the perturbed edge modes. The perturbation theory predicts that if the expectation value of the perturbation $\epsilon V$ is much smaller than $|E_1(k_y)-E_2(k_y)|$, the eigenvalues of $H(k_y)+\epsilon V(k_y)$ corresponding to $E_1(k_y)$ and $E_2(k_y)$ are $E'_1(k_y)=E_1(k_y)+\mathcal{O}(\epsilon)$ and $E'_2(k_y)=E_2(k_y)+\mathcal{O}(\epsilon)$. Then the change of the distance of the two edge eigenvalues is $\mathcal{O}(\epsilon)$. Since this change is much smaller than the original distance $|E_1(k_y)-E_2(k_y)|$, two edge modes are always nondegenerate in adding perturbations. Therefore, reconfiguration of the two edge modes are prohibited, and the edge modes remain connecting the upper and lower bulk bands, which implies the existence of gapless edge modes in the real part of the eigenenergy. 

\subsection{Spreading of the edge wave packet induced by the complex frequencies}
In the real-space simulation of the topological insulator laser (cf. Fig.~5 in the main text), the edge wave packet gradually spreads. To understand the origin of such spreading, we consider the one-dimensional wave packet described as 
\begin{equation}
 u(x,t) = \int dk \phi(k) e^{i(kx-\omega(k)t)}.
\end{equation}
By expanding the index around the wavenumber $k_0$, which maximizes $\rm{Im}\,\omega$, we obtain the following expression,
\begin{equation}
 u(x,t) \simeq e^{i(k_0 x-\omega(k_0)t)} \int dk \phi(k) e^{i(x-v_g t)(k-k_0)} e^{\frac{\partial^2 \rm{Im}\,\omega}{\partial k^2} (k-k_0)^2 t},
\end{equation}
where $v_g = \frac{\partial \rm{Re}\,\omega}{\partial k}$ is the group velocity of the wave packet. The last exponential represents the spreading wave packet, which does not appear in the real dispersion. To see the spreading clearly, we consider $\phi(k)=1$ and obtain the wave packet
\begin{eqnarray}
 u(x,t) = \sqrt{\frac{\pi}{At}} e^{i(k_0 x-\omega(k_0)t)} e^{-\frac{(x-v_g t)^2}{4At} },
\end{eqnarray}
where $A=-\partial^2 \rm{Im}\,\omega / \partial k^2 >0$. This implies that the width of the wave packet increases in proportional to the square root of time. 

We also calculate the group velocity and the spreading speed of the lasing wave packets described by the following edge effective Hamiltonian,
\begin{equation}
 H_{\rm edge}(k) = \left(
  \begin{array}{cc}
   E_0 +ak & i\beta \\
   i\beta' & E_0 -k
  \end{array}
  \right),\ \ a\neq 1.
\end{equation}
This Hamiltonian exhibits the eigenenergies $E=E_0 + [(a-1)k \pm \sqrt{(a+1)^2 k^2 -4\beta\beta'} ]/ 2$ and the imaginary parts take the maximum or minimum values at the wavenumber $k=0$. Therefore, the group velocity $v_g$ and the spreading speed $A$ are $v_g = (a-1)/2$ and $A = (a+1)^2/(4\sqrt{\beta\beta'})$.

\subsection{Two-layered Bernevig-Hughes-Zhang topological insulator laser}
\begin{figure}[t]
\includegraphics[width=80mm,bb=0 0 415 640,clip]{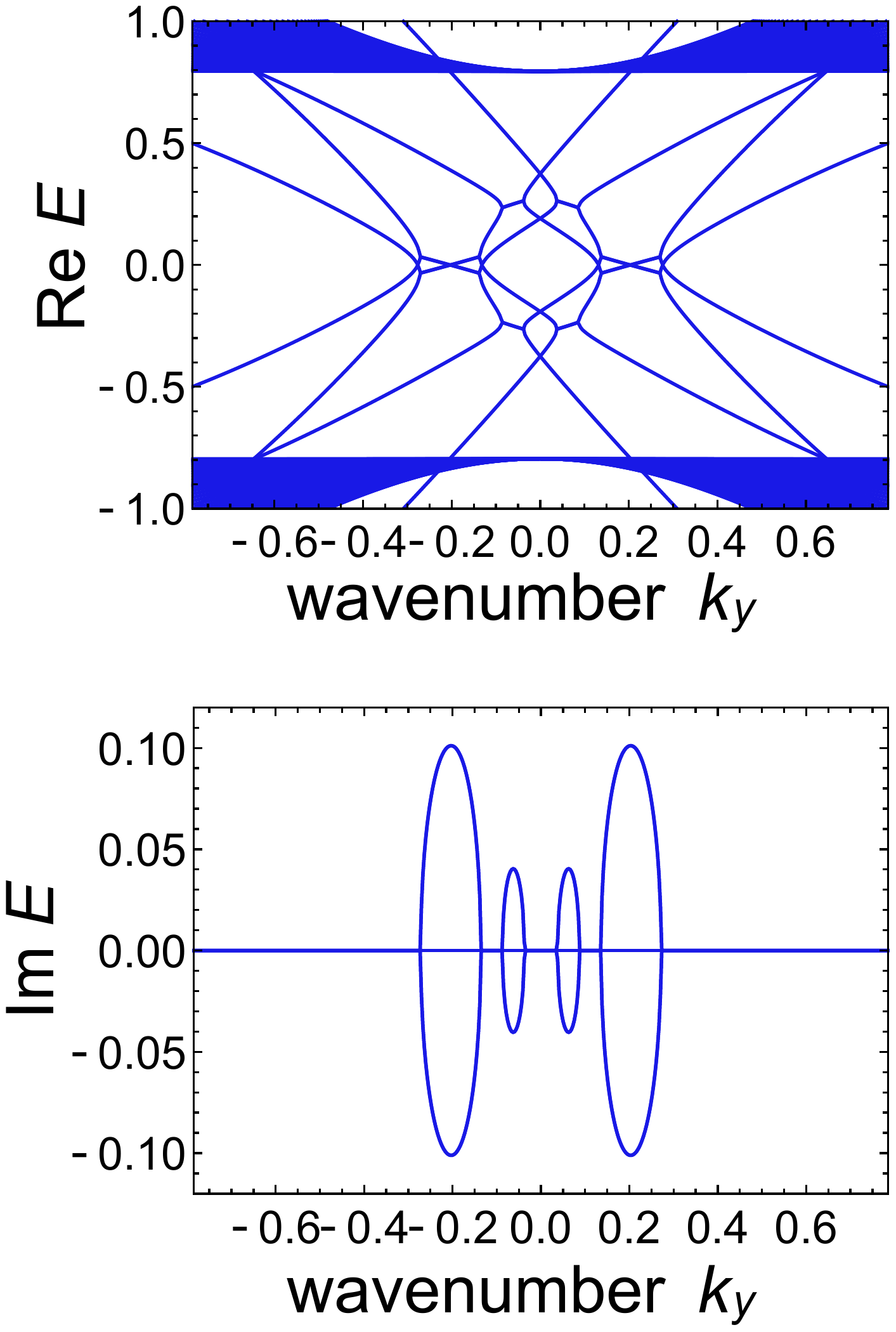}
\caption{\label{supplefig3}
Lasing edge modes in the modified two-layered Bernevig-Hughes-Zhang model. We arrange 50 sites in the $x$-direction with the open boundary and impose the periodic boundary condition in the $y$-direction. The parameters used are $u=-1$, $c=0.2$, $\beta=0.14$, and $\beta'=0.06$.
}
\end{figure}
\noindent
The two-layered Bernevig-Hughes-Zhang model (cf. Eq.~(1) and Fig.~1b in the main text) exhibits the nonzero imaginary part of the eigenenergies only in the exceptional edge modes as is the case for the topological insulator laser \cite{Harari2018,Bandres2018}. Thus, we can construct another topological insulator laser by modifying that model. The Hamiltonian for such a topological laser is described as
\begin{equation}
 H = \left(
  \begin{array}{cccc}
   2H_{\rm QWZ} & 2cI_2 & i \beta \sigma_x & 0 \\
   2cI_2 & 2H_{\rm QWZ} & 0 & i \beta' \sigma_x \\
    i \beta \sigma_x & 0 & H^{\ast}_{\rm QWZ} & cI_2 \\
   0 & i \beta' \sigma_x & cI_2 & H^{\ast}_{\rm QWZ} 
  \end{array}
  \right).
\end{equation}
As we constructed the topological insulator laser analyzed in the main text  (cf. Eq.~(5) and Fig.~4 in the main text), we modify the strength of the hopping in the two-layered QWZ model and couple the time-reversal two-layered QWZ model with a non-Hermitian coupling. Supplementary Figure~\ref{supplefig3} shows the edge band structure of this Hamiltonian. We can confirm the nonzero slope and the nonzero imaginary part of the energy in the energy dispersion corresponding to the lasing, mobile edge modes. We note that the model has the twice number of inner degrees of freedom compared to the topological insulator analyzed in the main text and thus exhibits more exceptional edge modes than in Fig.~4 in the main text.

\subsection{Exceptional edge modes in a continuum system}
To show that continuum Hamiltonians can also exhibit exceptional edge modes, we construct a continuum model with the Hamiltonian
\begin{equation}
H =   \left(
  \begin{array}{cc}
   H_0 & C \\
   -C^{\ast} & H_0^{\ast}
  \end{array}
  \right),
  \label{cont-toy}
\end{equation}
where $H_0$ is the continuum Hamiltonian for a Chern insulator with two gapless modes per edge:
\begin{equation}
H_0 =   \left(
  \begin{array}{cc}
   M-\beta \nabla^2 & a(-i\partial_x - \partial_y)^2 \\
   a(-i\partial_x + \partial_y)^2 & -M+\beta \nabla^2
  \end{array}
  \right),
\end{equation}
and $C$ is the non-Hermitian coupling:
\begin{equation}
C =   \left(
  \begin{array}{cc}
   0 & ib \\
   ib' & 0 
  \end{array}
  \right)
\end{equation}
with $a$, $b$, $b'$, $M$, and $\beta$ being real parameters. This model has the time-reversal symmetry in the same way as in the tight-binding model analyzed in the main text. By numerically diagonalizing the Hamiltonian, we obtain the edge band structure shown in Fig.~6 in the main text, where two gapless modes exist per edge, indicating the trivial bulk. However, those edge modes contain EPs and thus are stabilized against the disorder.

\subsection{Detail of the possible candidate of the proposed experimental setup of chiral active matter model}
\begin{figure}[t]
\includegraphics[width=160mm,bb=0 0 1110 370,clip]{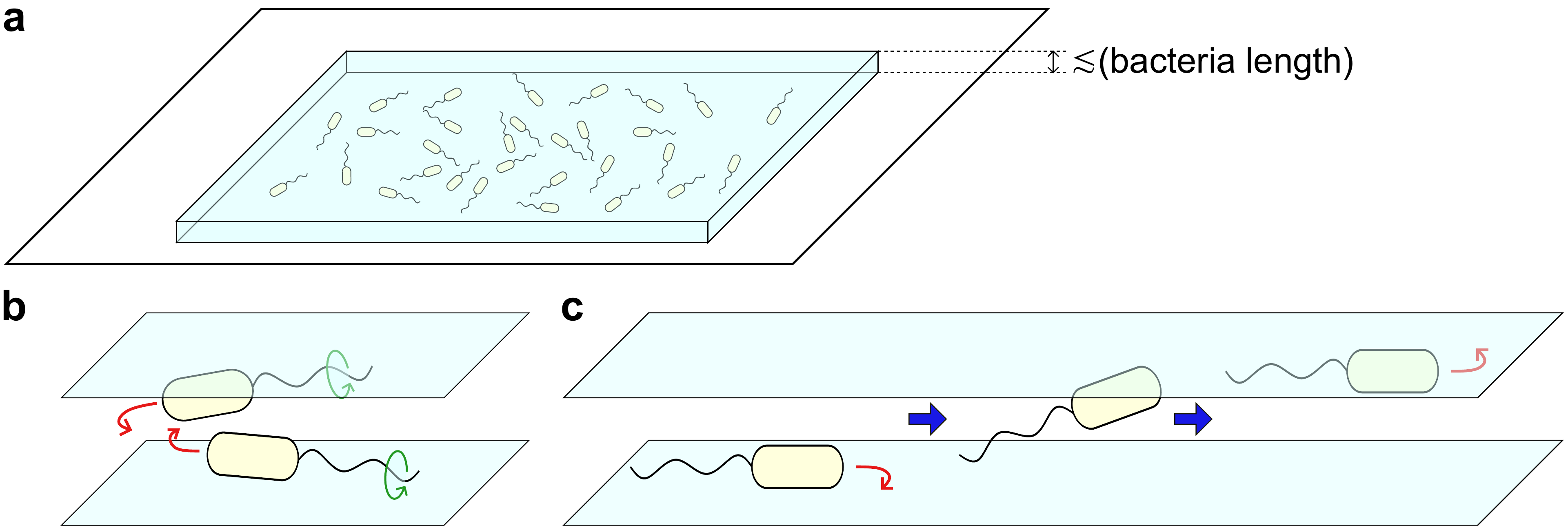}
\caption{\label{supplefig4}
Proposed experimental setup for the exceptional edge modes in chiral active matter. (a) Bacteria swimming between the upper and lower plates are considered. The distance between the two plates is set to be slightly shorter than the bacteria length. (b) Bacteria swimming near the lower (upper) plates shows the clockwise (counterclockwise) motion due to the rotation of flagella when we see them from the upper side of the system. By the collision between two bacteria near the upper and lower surfaces, interaction occurs between chiral active particles with opposite chirality. (c) The chirality of bacterial motion can flip by moving from the lower surface to the upper surface and vice versa.
}
\end{figure}
\noindent
We expect that the exceptional edge modes can appear in chiral active matter using bacteria. In more detail, we consider bacteria, such as {\it E.~coli}, swimming in the solution sandwiched by the upper and lower plates as shown in Supplementary Fig.~\ref{supplefig4}. We set the distance between the two plates about 1 $\mu$m, i.e., slightly shorter than the bacteria length to realize both the chirality flipping and the interaction between bacteria with different chiralities. Chirality of the bacterial motion occurs due to the rotation of flagella and their interaction with the plate \cite{DiLuzio2005,Li2008}. Therefore, the direction of motion depends on which surface the bacteria are close to and can be flipped by shuttling between the upper and lower surfaces. For the clear chirality of the bacterial motion, the distance of two plates must be enough large. Previous research \cite{Li2008} has shown that the rotational movement disappears if the distance between swimming bacteria and the surface plate becomes larger than a few tens percent of the length of the bacteria. Thus, we can expect that in the proposed system, one can observe the clear difference between the direction of rotation between the bacteria near the upper and lower surfaces.

On the other hand, aligning interaction between bacteria mainly originates from the collision of two bacteria. Since we set the distance between two plates shorter than the bacteria length, bacteria near the upper and lower surfaces can collide and interact with each other. One remaining problem is which type of interactions, nematic or polar interaction, dominates in the system. Previous research \cite{Bar2020} has revealed that if the repulsion between two self-propelled rods like bacteria is dominant compared to the self-propulsion, a polar cluster tends to appear. Since our proposal is based on chiral active matter with polar interaction, exceptional edge modes may be more realizable when we suppress the activity of bacteria by, for example, controlling the concentration of the solution.

In our analysis, we consider the situation that anti-polar interaction emerges in hydrodynamics of active matter (see the next section for details), while it is not obvious whether bacteria indeed exhibit such macroscopic anti-polar interaction or not in real experimental situations. As another candidate of chiral active matter that might show effective anti-polar interaction, we suggest the use of L-shaped active particles studied in previous research of chiral active matter \cite{Kummel2013}. L-shaped active particles can also lead to polar and/or anti-polar interaction and realize chirality flipping via the collision and the turnovers of the particles, thus possibly exhibiting exceptional edge modes.

\subsection{Derivation of the effective Hamiltonian for two-component chiral active matter}
We describe the detailed derivation of the effective Hamiltonian for our active matter model via the Boltzmann-Ginzburg-Landau approach \cite{Bertin2006,Peshkov2014}. We start with the particle model with polar (and anti-polar) interactions and rotational forces. We assume that the system is dilute such that three- and more-particle collisions are not significant. The dynamics of each particle is described as
\begin{eqnarray}
 \mathbf{r}_j(t+dt) &=& \mathbf{r}_j(t) + v_0 dt\mathbf{n}(\theta_j),\\
 \theta_j(t+dt) &=& \begin{cases}
  \theta_j(t) + \omega^i dt + \xi_j & \text{(no confliction)}, \\
  \Phi^{i,k}(\theta_j(t),\theta_l(t)) + \omega^i dt + \xi_j & \text{(when $j$th and $l$th particles conflict)},
 \end{cases}
\end{eqnarray}
where $\mathbf{r}_j$ and $\theta_j$ are the location and the direction of the $j$th particle and $i$ represents the direction of rotation. $\mathbf{n}(\theta_j)$ is the unit vector whose argument is $\theta_j$. $\omega^i$ creates the constant rotational force and $\xi_j$ is the Gaussian noise on the direction of each particle, which satisfies $\langle \xi_j(t) \rangle = 0$ and $\langle \xi_j(t) \xi_k(t') \rangle = T \delta_{jk} \delta(t-t')$ with $T$ being the effective temperature. $\Phi^{i,k}$ is the angle after the collision between the $i$- and $k$-rotating particles, which can take, e.g., the mean value of the angles of two colliding particles to describe the polar interaction. Also, we assume that the chirality $i$ flips occasionally at the rate $\gamma$. We note that if we consider $\omega^i=0$ and only one component, this model reduces to the Vicsek model \cite{Vicsek1995} except for the absence of the three- or more-particle confliction.

From the stochastic particle model, we derive the following Boltzmann equation,
\eqn{
&{}& \partial_t f^i + v_0 \mathbf{n} \cdot \nabla f^i + \omega^i \partial_{\theta} f^i \nonumber\\ 
&=& D_0 \Delta f^i + D_1 g^{\alpha\beta} \partial_{\alpha}\partial_{\beta} f^i + I_{{\rm dif}} [f^i] + I_{{\rm col}}^{{\rm hom}} [f^i] + \sum_{j\neq i} I_{{\rm col}}^{{\rm het}} [f^i,f^j] + \sum_{j\neq i}\gamma(f^j-f^i),
}
where $f^i$ is the one-particle distribution function for the $i$-rotating particles and 
\begin{eqnarray}
 I_{{\rm dif}} [f^i] &=& -\lambda f^i + \lambda \int d\theta' \int d\eta P_{\sigma}(\eta) \delta_{m\pi} (\theta'+\eta-\theta) f^i(\mathbf{r},\theta',t),\\
 I_{{\rm col}}^{{\rm hom}} [f^i] &=& -f^i(\mathbf{r},\theta,t) \int d\theta' K(\theta'-\theta) f^i(\mathbf{r},\theta',t) \nonumber\\
 &{}& + \int d\theta_1 \int d\theta_2 \int d\eta P_{\sigma}(\eta) K^i(\theta_2-\theta_1) f^i(\mathbf{r},\theta_1,t) f^i(\mathbf{r},\theta_2,t) \delta_{m\pi} (\Phi^{i,i}(\theta_1,\theta_2)+\eta-\theta), \nonumber \\
 &{}&\\
 I_{{\rm col}}^{{\rm het}} [f^i,f^j] &=& -f^i(\mathbf{r},\theta,t) \int d\theta' K(\theta'-\theta) f^j(\mathbf{r},\theta',t) \nonumber\\
 &{}& + \int d\theta_1 \int d\theta_2 \int d\eta P_{\sigma}(\eta) K^{i,j}(\theta_2-\theta_1) f^i(\mathbf{r},\theta_1,t) f^j(\mathbf{r},\theta_2,t) \delta_{m\pi} (\Phi^{i,j}(\theta_1,\theta_2)+\eta-\theta), \nonumber \\
 &{}&
\end{eqnarray}
are the self-diffusion integral, the collision integral between the same species, and that between the different species for each. $P(\eta)$ is the noise distribution, which is supposed to be Gaussian in the present setup, and $K(\theta_2-\theta_1)$ represents the collision kernel. Here, we utilize the molecular chaos hypothesis, which assumes that the two-body distribution function can be described as the product of the one-body distribution functions.

To derive hydrodynamic equations, we conduct the Fourier transformation of the obtained Boltzmann equation. We consider the following Fourier(-like) components:
\eqn{
 f_k^i (\mathbf{r},t) &=& \int_{-\pi}^{\pi} f^i (\mathbf{r},\theta,t) e^{ik\theta} d\theta, \\
 P_k (\mathbf{r},t) &=& \int_{-\pi}^{\pi} P (\eta) e^{ik\eta} d\eta, \\
 I_{k,q}^i (\mathbf{r},t) &=& \frac{1}{2\pi} \int_{-\pi}^{\pi} K^i (\Delta) e^{-iq\Delta+ik\Phi^{i,i}(0,\Delta)} d\Delta, \\
 I_{k,q}^{i,j} (\mathbf{r},t) &=& \frac{1}{2\pi} \int_{-\pi}^{\pi} K^{i,j} (\Delta) e^{-iq\Delta+ik\Phi^{i,j}(0,\Delta)} d\Delta.
}
It is noteworthy that one can calculate $P_k$, $I_{k,q}^i$, and $I_{k,q}^{i,j}$ from the microscopic parameters. Performing the Fourier transformation, we obtain the following equation:
\begin{eqnarray}
 &{}& \frac{\partial f_k^i}{\partial t} + \frac{v_0}{2} (\nabla f_{k-1}^i + \nabla^{\ast} f_{k+1}^i)-ik \omega^i f_k^i \nonumber\\
 &=& -\lambda(1-P_k)f_k^i + D_0 \Delta f_k^i + \frac{D_1}{4} (\nabla^2 f_{k-2}^i + (\nabla^{\ast})^2 f_{k+2}^i) \nonumber\\ 
 &{}& + \sum_{q=-\infty}^{\infty} (P_k I_{k,q}^i - I_{0,q}^i ) f_q^i f_{k-q}^i + \sum_{j\neq i} \sum_{q=-\infty}^{\infty} (P_k I_{k,q}^{i,j} - I_{0,q}^{i,j} ) f_q^i f_{k-q}^j + \sum_{j\neq i} \gamma (f_k^j - f_k^i),
 \label{FourierBoltzmann}
\end{eqnarray}
where $\nabla = \partial_x + i\partial_y$ and $\nabla^{\ast} = \partial_x - i\partial_y$. By definition, $f_0^i$ and $({\rm Re}f_1^i,{\rm Im}f_1^i)^T$ are equivalent to the density and the velocity field, respectively. With a little algebra, we obtain the equations for $k=0,1,2$,
\begin{eqnarray}
&{}& \frac{\partial\rho^i}{\partial t} + v_0 {\rm Re}(\nabla^{\ast} f_1^i) = D_0 \Delta \rho^i + \frac{D_1}{2} {\rm Re}(\nabla^{\ast 2} f_2^i)+\sum_{j\neq i} \gamma(\rho^j-\rho^i),
 \label{FourierBoltzmann0}\\
 &{}& \frac{\partial f_1^i}{\partial t} +\frac{v_0}{2} (\nabla \rho^i + \nabla^{\ast} f_2^i) - i\omega^i f_1^i \nonumber\\
 &=& [\lambda(P_1-1) + (P_1 I_{1,0}^i - I_{0,0}^i + P_1 I_{1,1}^i - I_{0,1}^i ) \rho^i ] f_1^i + D_0 \Delta f_1^i + \frac{D_1}{4} \nabla^2 (f_1^i)^{\ast} \nonumber\\
 &{}&  + \sum_{j\neq i} (P_1 I_{1,1}^{i,j}-I_{0,1}^{i,j}) \rho^j f_1^i + \sum_{j\neq i} (P_1 I_{1,1}^{i,j}-I_{0,1}^{i,j}) \rho^i f_1^j \nonumber\\
 &{}& + (P_1 I_{1,2}^i - I_{0,2}^i + P_1 I_{1,-1}^i - I_{0,-1}^i) f_2^i (f_1^i)^{\ast} \nonumber\\
 &{}& + \sum_{j\neq i} (P_1 I_{1,-1}^{i,j} - I_{0,-1}^{i,j}) f_2^j (f_1^i)^{\ast} + \sum_{j\neq i} (P_1 I_{1,2}^{i,j} - I_{0,2}^{i,j}) f_2^i (f_1^j)^{\ast}+\sum_{j\neq i} \gamma(f_1^j-f_1^i),
 \label{FourierBoltzmann1}\\
 &{}& \frac{\partial f_2^i}{\partial t} + \frac{v_0}{2} \nabla f_2^i - 2i\omega^i f_2^i \nonumber\\
 &=& \left(A^i + B^i \rho^i + \sum_{j\neq i} C^{i,j} \rho^j \right) f_2^i + \frac{D_1}{4} \nabla^2 \rho^i + (P_2 I_{2,1}^i - I_{0,1}^i) (f_1^i)^2 \ \nonumber\\
 &{}&  + \sum_{j\neq i} (P_2 I_{2,1}^{i,j} - I_{0,1}^{i,j}) f_1^i f_1^j + \sum_{j\neq i} (P_2 I_{2,0}^{i,j} - I_{0,0}^{i,j}) \rho^i f_2^j+\sum_{j\neq i} \gamma(f_2^j-f_2^i),
 \label{FourierBoltzmann2}
\end{eqnarray}
where we omit the high-frequency terms, $f_{k'}^i$ ($k'\geq3$), which are irrelevant in the following discussion. Since we consider the polar active matter (that can have anti-polar interaction only between oppositely rotating particles), $f_1^i$ becomes the leading order term. We have to balance the following terms in Supplementary Eqs.~\eqref{FourierBoltzmann0}, \eqref{FourierBoltzmann1}, \eqref{FourierBoltzmann2},
\begin{equation}
 \partial_t \rho^i \sim {\rm Re} \nabla^{\ast} f_1^i ,\ \ \partial_t f_1^i \sim \nabla \rho^i,\ \ f_2^i \sim (f_1^i)^2,\ \ (f_1^i)^{\ast} f_2^i \sim \Delta f_1^i.
\end{equation}
Therefore, we obtain the following scaling relations,
\begin{equation}
 f_1^i \sim \epsilon,\ \ f_2^i \sim \epsilon^2,\ \ \rho^i-\rho^i_{\rm ss} \sim \epsilon,\ \ \nabla\sim\epsilon,\ \ \partial_t \sim \epsilon.
\end{equation}
By considering the terms with the order $\epsilon^2$, we can confirm that the $f_2^i$ can be described as a linear combination of $\nabla f_1^j$, $( f_1^j)^2$, and $f_1^j f_1^l$. By substituting this, we finally obtain the hydrodynamic equations,
\begin{eqnarray}
 \partial_t \rho^i + v_0 \nabla \cdot \mathbf{p}^i &=& D_0 \Delta \rho^i + \sum_j (\gamma \rho^j - \gamma \rho^i), \\ \label{CAM-fluid1}
 \partial_t \mathbf{p}^i + \lambda (\mathbf{p}^i \cdot \nabla) \mathbf{p}^i &=& \boldsymbol{\omega}^i \times \mathbf{p}^i + (a^i(\rho) + b^i(\mathbf{p})) \mathbf{p}^i  + \sum_{j\neq i} (c^{i,j}(\rho) + d^{i,j}(\mathbf{p})) \mathbf{p}^j  \nonumber\\
 &{}& - \mu^i \nabla \rho^i + D_0 \Delta \mathbf{p}^i + D_{\rm odd} \Delta \mathbf{p}^{i\ast} + ({\rm higher\mathchar`-order\ terms}) \label{CAM-fluid2},
\end{eqnarray}
where $i$ represents the chirality of active matter, i.e., the anticlockwise ($i=1$) and clockwise ($i=2$) moving direction. $\rho^i(\mathbf{r},t)$ is the density field of active matter and $\mathbf{p}^i(\mathbf{r},t)=\rho^i(\mathbf{r},t)\mathbf{v}^i(\mathbf{r},t)/v^i_{\rm ss}$ is the momentum field divided by the steady-state velocity, where $\mathbf{v}^i(\mathbf{r},t)$ is the local average of velocities of self-propelled particles. Here, we omit some derivative terms and some higher-order terms with respect to $\epsilon$. However, we remain the higher-order term including $\mathbf{p}^{i\ast}=(-p^i_y,p^i_x)^T$ to make the effective Hamiltonian compact (shown to be important for defining the topological invariant in continuum space \cite{Bal2019}). This term is called odd viscosity and has been derived in the hydrodynamic equations of chiral active matter in previous research \cite{Banerjee2017}. We note that these equations are similar to the Toner-Tu equations \cite{Toner1995}, hydrodynamic equations for one-component polar active matter. However, unlike the Toner-Tu equations, they contain a Coriolis-force-like term, $\boldsymbol{\omega}^i \times \mathbf{p}^i$, and momentum-coupling terms, $(c^{i,j}(\rho) + d^{i,j}(\mathbf{p})) \mathbf{p}^j$. The origins of the Coriolis-force-like term and the momentum-coupling terms are the chirality and the polar or anti-polar interaction between particles with the different chiralities. We note that there still remains the possibility of inconsistency between the polar or anti-polar interaction in the microscopic stochastic description and the hydrodynamic equations, because in the parameter region of the unordered phase, polar (or anti-polar) interaction does not lead to collective motion and instead can possibly enhance the anti-ordering with a help of the rotational motion of active matter. In the discussion below, we focus on the case that there is (effectively) anti-polar interaction in the hydrodynamic description between particles with opposite chiralities.

Linearizing the above hydrodynamic equations around a steady-state solution, we derive differential equations for the fluctuations of density and velocity fields, 
\begin{eqnarray}
 \partial_t \delta \rho^i + \mathbf{v}_{{\rm ss}}^i \cdot \nabla \delta \rho^i &=& -\rho_{\rm ss}^i \nabla \cdot \delta \mathbf{v}^i + \sum_{j\neq i} \gamma (\rho^j - \rho^i), \\
 \partial_t \delta \mathbf{v}^i + \lambda (\mathbf{v}_{{\rm ss}}^i \cdot \nabla) \delta \mathbf{v}^i &=& \boldsymbol{\omega}^i \times \delta \mathbf{v}^i + \alpha^i \delta \mathbf{v}^i + \sum_{j\neq i} \beta^{i,j} \delta \mathbf{v}^j - \mu^i \nabla \delta \rho^i + \nu^o \Delta \delta \mathbf{v}^{i\ast},
\end{eqnarray}
where $\rho^i_{\rm ss}$ and $\mathbf{v}^i_{\rm ss}$ represent the steady-state values of the density and the velocity field, respectively, and  $\delta\rho^i({\bf r},t) = \rho^i({\bf r},t) - \rho^i_{\rm ss}({\bf r})$ and $\delta\mathbf{v}^i({\bf r},t)=\mathbf{v}^i({\bf r},t)-\mathbf{v}^i_{\rm ss}({\bf r})$ are their fluctuations. While the above equations describe the linear dynamics around arbitrary steady-state solutions, here we consider the nonordered steady state, $\rho^1_{\rm ss}= \rho^2_{\rm ss}\equiv\rho_{\rm ss}$, $\mathbf{v}^i_{\rm ss}=0$, and the symmetric or antisymmetric parameters, $c^{1,2}(\rho_{\rm ss}) + d^{1,2}(\mathbf{p}_{\rm ss}) = c^{2,1}(\rho_{\rm ss}) + d^{2,1}(\mathbf{p}_{\rm ss}) = a^i(\rho_{\rm ss}) + b^i(\mathbf{p}_{\rm ss}) \equiv -\beta$, $\omega^1 = -\omega^2 \equiv \omega_0$, $\rho^1_{\rm ss} = \rho^2_{\rm ss} \equiv \rho_0$, $\gamma^{1,2}=\gamma^{2,1}\equiv\gamma$. By nondimesionalizing the equations, we finally obtain the following linearized equations in the frequency domain, 
\begin{equation}
  \omega
  \left(
  \begin{array}{c}
   \delta \tilde{\rho}^1 \\
   \delta \tilde{v}_x^1 \\
   \delta \tilde{v}_y^1 \\
   \delta \tilde{\rho}^2 \\
   -\delta \tilde{v}_x^2 \\
   -\delta \tilde{v}_y^2
  \end{array}
  \right) = H 
  \left(
  \begin{array}{c}
   \delta \tilde{\rho}^1 \\
   \delta \tilde{v}_x^1 \\
   \delta \tilde{v}_y^1 \\
   \delta \tilde{\rho}^2 \\
   -\delta \tilde{v}_x^2 \\
   -\delta \tilde{v}_y^2
  \end{array}
  \right) \label{CAM-linear}
\end{equation}
with $H$ being the effective Hamiltonian defined as
\begin{equation}
H =   \left(
  \begin{array}{cc}
   H_0 +A & C \\
   -C^{\ast} & H_0^{\ast} +A
  \end{array}
  \right) \label{CAM-Hamiltonian},
\end{equation}
\begin{equation}
H_0 =   \left(
  \begin{array}{ccc}
   0 & -i \partial_x & -i \partial_y \\
   -i\partial_x & 0 & -i(\omega_0+\nu^o \Delta) \\
   -i \partial_y & i(\omega_0+\nu^o \Delta) & 0
  \end{array}
  \right) \label{CAM-H0},
\end{equation}
\begin{equation}
A =   \left(
  \begin{array}{ccc}
   -i\gamma & 0 & 0 \\
   0 & -i\beta & 0 \\
   0 & 0 & -i\beta
  \end{array}
  \right) \label{CAM-diagonal}.
\end{equation}
\begin{equation}
C =   \left(
  \begin{array}{ccc}
   i\gamma & 0 & 0 \\
   0 & i\beta & 0 \\
   0 & 0 & i\beta
  \end{array}
  \right) \label{CAM-coupling},
\end{equation}
where $\delta\tilde{\rho}^i$ and $\delta\tilde{v}_{x,y}^i$ represent the Fourier components of the nondimensionalized fluctuation of the density and the velocity field, respectively, and all the parameters are also nondimensionalized.
We note that the Hamiltonian $H_0$ represents the effective Hamiltonian for topological fluid, which can be realized by utilizing chiral active matter \cite{Souslov2019}. Our Hamiltonian is no longer time-reversal symmetric. Since we combine the two topological systems with the opposite Chern numbers, the edge modes in Fig.~7b are not protected by the bulk topology. On the other hand, the coupling term $C$ is non-Hermitian and thus creates EPs in the edge bands that protect the gapless modes. For numerical calculation, we use the parameters $\omega=1$, $\nu=0.5$, $\gamma=0.3$, and $\beta=0.5$.

While we have considered the two-valued rotational force $\omega$, the distribution of rotational velocities must be continuous in practice. However, if we can separate moving particles into two groups with clockwise and anticlockwise rotations, the coarse-grained dynamics of chiral active matter should not be changed. We can model such a situation by considering the following stochastic process of $\omega$,
\begin{equation}
\frac{d\omega}{dt} = -\frac{\partial U}{\partial \omega} + \xi_{\omega},
\end{equation}
where $U$ and $\xi_{\omega}$ represent the effective potential and the Gaussian noise, respectively. If we consider a double-well potential $U$ that has minima at $\omega=\pm\omega_0$, the particles can be divided by their directions of rotations. We can derive the Boltzmann equation,
\eqn{
&{}& \partial_t f + v_0 \mathbf{n} \cdot \nabla f + \omega \partial_{\theta} f \nonumber\\ 
&=& D_0 \Delta f + D_1 g^{\alpha\beta} \partial_{\alpha}\partial_{\beta} f + I_{{\rm dif}} [f] + I_{{\rm col}} [f] + \partial_{\omega} \left[ \partial_{\omega}U + D_{\omega} \partial_{\omega} \right] f,
}
where $f=f(\mathbf{r},\theta,\omega,t)$ is the one-particle distribution function, and $I_{{\rm dif}} [f]$ and $I_{{\rm col}} [f]$ are the self-diffusion integral and the collision integral for each. By integrating the equation for $\omega<0$ ($\omega>0$), we can derive the time evolution of the one-particle distribution function for clockwise (anticlockwise) particles and check that the obtained equations are equivalent to the equations with the two-valued rotational force.

\subsection{Detail of the band structure of the active matter model}
\begin{figure}[t]
\includegraphics[width=160mm,bb=0 0 800 200,clip]{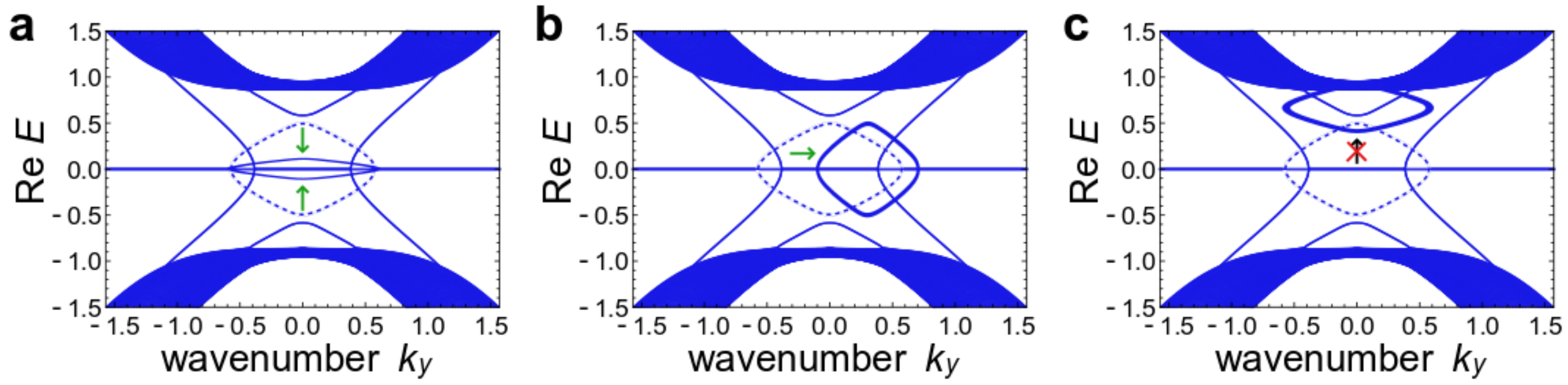}
\caption{\label{supplefig5}
Allowed and prohibited modifications of the dispersion relation in the chiral active matter model. (a)-(c) Each figure shows what modification of the dispersion relation with a diamond shape in Fig.~7 in the main text is allowed (the green arrows) or prohibited (the black arrow with the red cross). (a) We can push the edge dispersion into the bulk dispersion around the ${\rm Re}\,E=0$ axis. (b) Moving the edge dispersion to the right is also possible. In this case, the apparent crossing disappears. (c) Pushing up the edge dispersion to the upper bulk dispersion is prohibited.
}
\end{figure}
\noindent
Here, we discuss the meaning of the crossings around the ${\rm Re}\,E=0$ axis, which is not depicted as the red circle in Fig.~7 in the main text and what modifications of the edge dispersion are allowed. The crossings are the point where the unprotected edge dispersions with a diamond shape appear from the bulk dispersions around ${\rm Re}\,E=0$. Therefore, it is impossible to separate the unstable edge dispersions with a diamond shape and the bulk dispersions around ${\rm Re}\,E=0$. This constraint leads to the prohibition of moving the unstable edge dispersions to the upper or lower bulk dispersions as shown in Supplementary Fig.~\ref{supplefig5}c. On the other hand, these edge dispersions are not protected by the topology or the EP, we can push them into the bulk dispersions around ${\rm Re}\,E=0$ depicted in Supplementary Fig.~\ref{supplefig5}a. Besides, the crossings in the bulk gap are not true degeneracies since the edge bands have different imaginary parts of the eigenfrequencies and thus can be removed by moving the unstable edge dispersions to the right or left as demonstrated in Supplementary Fig.~\ref{supplefig5}b. 

\subsection{Bernard-LeClair symmetry class of our models}
\begin{figure}[t]
\includegraphics[width=80mm,bb=0 0 390 290,clip]{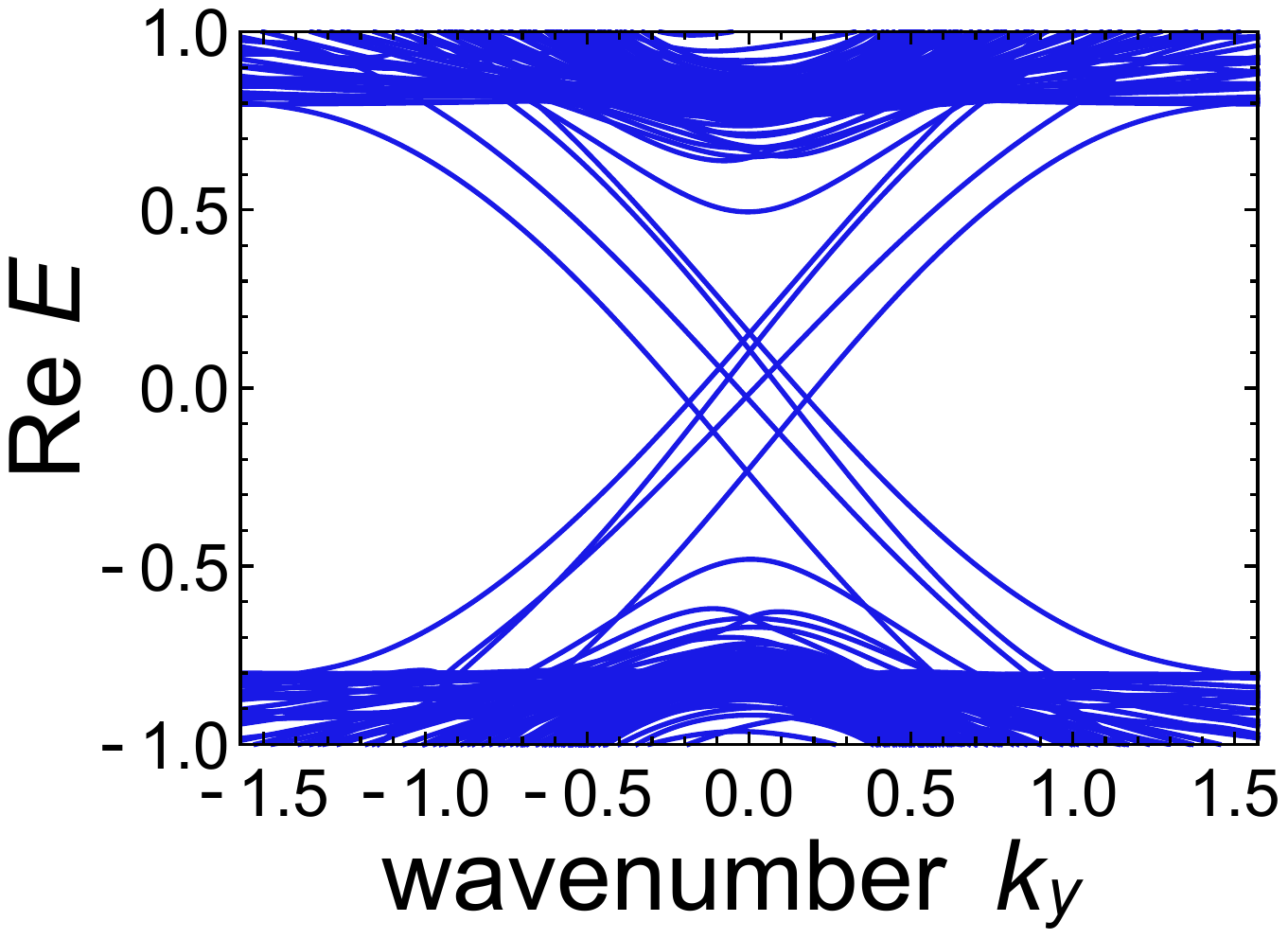}
\caption{\label{supplefig6}
Robust gapless edge modes in the $S_+$AII class. We calculate the edge band structure of the two-layered non-Hermitian Bernevig-Hughes-Zhang model with the on-site imaginary potential and the complex noise in the non-Hermitian coupling and without the Hermitian coupling. We align 50 sites in the $x$ direction with the open boundary and impose the periodic boundary condition in the $y$ direction. The parameters used are $u=-1$, $c=0.2$, $\beta=0.14$, $\beta'=0.06$, $\gamma=0$, and $g=1.1$. The noise width for the real noise in the non-Hermitian coupling is $W=1.0$.
}
\end{figure}
\noindent
In Hermitian systems, three types of $\mathbb{Z}_2$ symmetries (i.e., the time-reversal symmetry, the particle-hole symmetry, and the chiral (sublattice) symmetry) play an important role in the classification of topological band structures \cite{Altland1997,Hasan2010}. The time-reversal (particle-hole, chiral) symmetry means that there exists a unitary operator $T$ ($C$, $\Gamma$) which satisfies $TH(\mathbf{k})T^{-1} = H(-\mathbf{k})^{\ast}$ ($CH(\mathbf{k})C^{-1} = -H(-\mathbf{k})^T$, $\Gamma H(\mathbf{k})\Gamma^{-1} = -H(\mathbf{k})^{\dagger}$) and $TT^{\ast}=\pm 1$ ($CC^{\ast}=\pm 1$, $\Gamma^2=1$), where $H(\mathbf{k})$ is the Bloch Hamiltonian. Because of $\Gamma = CT$, if there are two of these symmetries, the system also exhibits the other $\mathbb{Z}_2$ symmetry. Therefore, we can classify Hermitian Hamiltonians into 10 Altland-Zirnbauer (AZ) classes \cite{Hasan2010} concerning the $\mathbb{Z}_2$ symmetries. 

Recent researches \cite{Gong2018,Zhou2019,Kawabata2019} have extended the $\mathbb{Z}_2$ symmetries and the AZ symmetry classes to non-Hermitian systems. We need to reconsider the discrepancy between $H$ and $H^{\dagger}$ and thus distinguish two types of each $\mathbb{Z}_2$ symmetry, i.e., $TH(\mathbf{k})T^{-1} = H(-\mathbf{k})^{\ast}$ and $TH(\mathbf{k})T^{-1} = H(-\mathbf{k})^{T}$ for the time-reversal symmetry, $CH(\mathbf{k})C^{-1} = -H(-\mathbf{k})^T$ and $CH(\mathbf{k})C^{-1} = -H(-\mathbf{k})^{\ast}$ for the particle-hole symmetry, $\Gamma H(\mathbf{k})\Gamma^{-1} = -H(\mathbf{k})^{\dagger}$ and $\Gamma H(\mathbf{k})\Gamma^{-1} = -H(\mathbf{k})$ for the chiral (sublattice) symmetry. In a previous paper \cite{Kawabata2019}, the symmetry $\Gamma H(\mathbf{k})\Gamma^{-1} = -H(\mathbf{k})^{\dagger}$ is called the chiral symmetry, and $\Gamma H(\mathbf{k})\Gamma^{-1} = -H(\mathbf{k})$ is called the sublattice symmetry, and the two other $\mathbb{Z}_2$ symmetries are distinguished by adding the dagger for the latter definition. For the topological classification, it is evident that the daggered time-reversal symmetry (the daggered particle-hole symmetry) coincides with the particle-hole symmetry (the time-reversal symmetry). By utilizing these extended $\mathbb{Z}_2$ symmetries, one can obtain the 38 Bernard-LeClair (BL) classes \cite{Zhou2019,Kawabata2019,Bernard2002a,Bernard2002b}. In non-Hermitian systems, the chiral symmetry and the sublattice symmetry are independent of each other, and thus increase the number of the symmetry classes. In particular, whether the sublattice-symmetry operator commutes or anticommutes with each of the other $\mathbb{Z}_2$ operators plays an important role in the classification.

The periodic table for topological insulators predicts the existence or absence of topological invariants for the Bloch Hamiltonian in each AZ class. To define the topological index in non-Hermitian Hamiltonians, we must reconsider the energy gap. In Hermitian systems, the eigenenergies distribute on the one-dimensional energy axis and thus have only point-like gaps in their spectra. On the contrary, in non-Hermitian systems, the eigenenergies can be complex and thus distribute on the two-dimensional space, which leads to the variety of the definitions of the energy gaps \cite{Gong2018}. Especially, the periodic table for non-Hermitian Hamiltonians deals with line gaps and point gaps. If we can draw a line that separates complex energy spectra, the band structures have the line gap. On the other hand, energy spectra can encircle a point in the complex energy space, which never occurs in Hermitian Hamiltonians. Such a point is called a point gap and leads to topological invariants unique to non-Hermitian systems. Besides, to properly take the constraint of the symmetry into account, we need to consider both the line gaps parallel to the real and imaginary axes of the complex energy space for some BL classes. The periodic table indicates what type of topological invariants appear for each kind of energy gap in the Hamiltonian belonging to each non-Hermitian BL class. 

One can discuss the symmetries and the topological invariants in the Hamiltonians of our models. Here, we focus on the lattice model with the on-site imaginary potential (cf. Fig.~3d in the main text) and without the Hermitian coupling $\gamma\sigma_x \otimes \sigma_y \otimes \sigma_x$. As in Fig.~3d in the main text, this model exhibits robust gapless edge modes shown in Supplementary Fig.~\ref{supplefig6}. However, these gapless edge modes are not characterized by the topological invariant predicted from the periodic table \cite{Gong2018,Zhou2019,Kawabata2019}. The Hamiltonian is described as follows in terms of the Pauli matrices $\sigma_i$,
\begin{eqnarray}
 H (\mathbf{k}) &=& (u+\cos k_x + \cos k_y) I_2 \otimes I_2 \otimes \sigma_z + \sin k_y I_2 \otimes I_2 \otimes \sigma_y + \sin k_x \sigma_z \otimes I_2 \otimes \sigma_x \nonumber\\
 &{}& + cI_2 \otimes \sigma_x \otimes I_2 + i\frac{\beta+\beta'}{2} \sigma_x \otimes I_2 \otimes \sigma_x + i\frac{\beta-\beta'}{2} \sigma_x \otimes \sigma_z \otimes \sigma_x + ig\sigma_z \otimes I_2\otimes I_2.
\end{eqnarray}
From this expression and the anticommutation relations of the Pauli matrices, we can confirm that the following unitary operators represent the $\mathbb{Z}_2$ symmetries of the Hamiltonian:
\begin{eqnarray}
T &=& (i\sigma_y) \otimes I_2 \otimes I_2,\ \ T H(\mathbf{k}) T^{-1} = H^{\ast} (-\mathbf{k})\ \ \ \ \ ({\rm time\ reversal\ symmetry}),\\
S &=& (i\sigma_y) \otimes \sigma_z \otimes \sigma_x ,\ \ S H(\mathbf{k}) S^{-1} = -H (\mathbf{k})\ \ \ ({\rm sublattice\ symmetry}).
\end{eqnarray}
Furthermore, these operators satisfy $TT^{\ast}=-1$ and $[T,S]=0$. Therefore, we conclude that this Hamiltonian belongs to the $S_+$AII class according to a proposed periodic table \cite{Kawabata2019} and thus should have only the trivial index. However, as shown in Supplementary~Fig.~\ref{supplefig6}, this Hamiltonian can exhibit robust gapless edge modes, which cannot be captured by the periodic table, thus demonstrating the breakdown of the bulk-edge correspondence. While one can readily check that the other Hamiltonians can have nontrivial bulk indices, exceptional edge modes in those models also can emerge independently of the bulk topological invariants. Thus, the predictions from the periodic table are irrelevant to the presence or absence of robust gapless edge modes proposed in the present work.

\subsection{$\mathbb{Z}_2$ symmetries in our models and their roles in the protection of exceptional edge modes}
\begin{figure}[b]
\includegraphics[width=80mm,bb=0 0 510 410,clip]{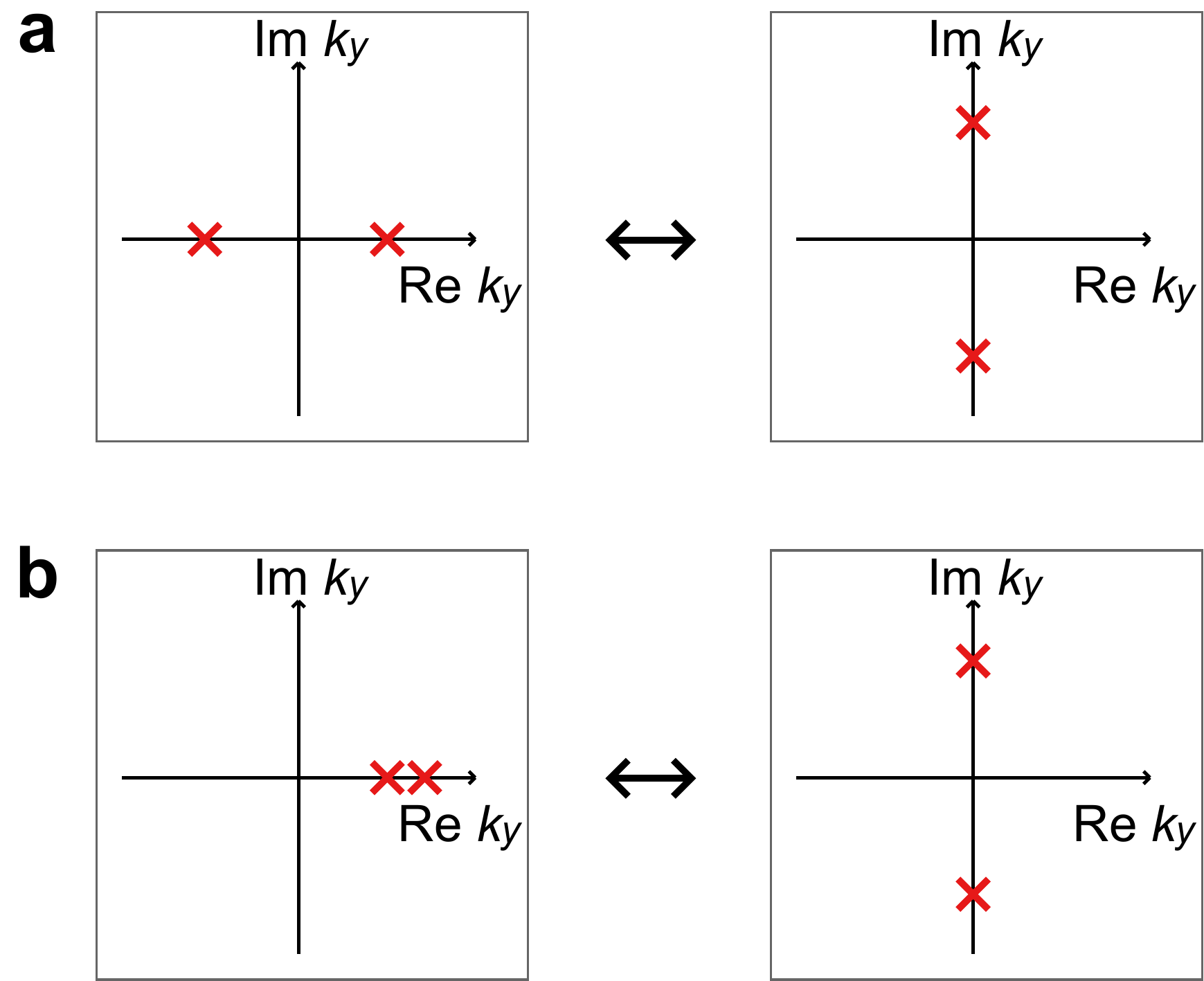}
\caption{\label{supplefig7}
Eigenvalues of edge modes around exceptional points. (a)-(b) Eigenvalues of edge modes around the EP are presented. (a) The eigenvalues of the two edge modes localized at the same side are changed from a pair of opposite real values to complex conjugates around the EP, for example in Fig.~1b around the ${\rm Re}\,E=0$ axis. (b) The eigenvalues of the edge modes localized at the opposite sides have the same eigenvalues before crossing the EP. The edge modes considered here are converted into each other by applying the conventional $PT$, $CP$, or modified pseudo-Hermitian operation.
}
\end{figure}
\noindent
Exceptional points in exceptional edge modes can be protected by the topology associated with the branchpoint singularity and the symmetries, such as the $PT$ symmetry, the $CP$ symmetry, the chiral symmetry, and the pseudo-Hermiticity \cite{Kawabata2019b} as discussed in the main text. To confirm this point, we describe the $\mathbb{Z}_2$ symmetries of our models related to the protection of EPs. The symmetry in the bulk system should agree with that in the edge and thus should determine whether the EPs can be topologically protected or not. However, we find that the conventional $PT$, $CP$ symmetries, and the modified pseudo-Hermiticity play no roles in determining the protection of exceptional edge modes, as discussed in the following paragraph.

First, we confirm the symmetries of the two-layered non-Hermitian Bernevig-Hughes-Zhang model. It has the $PT$ symmetry and the pseudo-Hermiticity defined as,
\begin{eqnarray}
 PT &=& \sigma_x \otimes I_2 \otimes \sigma_z,\ PTH(\mathbf{k})(PT)^{-1}=H^{\ast}(\mathbf{k}),\\
 \Gamma &=& \sigma_x \otimes \sigma_z \otimes \sigma_x, \ \Gamma H(\mathbf{k})\Gamma^{-1}=-H^{\dagger}(\mathbf{k}).
\end{eqnarray}
Furthermore, we can also define the following modified $PT$ symmetry:
\begin{eqnarray}
 P'T &=& I_2 \otimes I_2 \otimes \sigma_z, \ P'TH(k_x,k_y)(P'T)^{-1}=H^{\ast}(-k_x,k_y),\\
 P''T &=& \sigma_z \otimes I_2 \otimes I_2, \ P''TH(k_x,k_y)(P''T)^{-1}=H^{\ast}(k_x,-k_y).
\end{eqnarray}
The model without the Hermitian coupling ($\gamma=0$) also has the conventional and modified $CP$ symmetry and pseudo-Hermiticity defined as,
\begin{eqnarray}
 CP &=& \sigma_z \otimes \sigma_z \otimes \sigma_y, \ CPH(\mathbf{k})(CP)^{-1}=-H^{\ast}(\mathbf{k}),\\
 CP' &=& \sigma_y \otimes \sigma_z \otimes \sigma_y, \ CP'H(k_x,k_y)(CP')^{-1}=-H^{\ast}(-k_x,k_y),\\
 CP'' &=& \sigma_x \otimes \sigma_z \otimes \sigma_x, \ CP''H(k_x,k_y)(CP'')^{-1}=-H^{\ast}(k_x,-k_y),\\
 \eta &=& \sigma_z \otimes I_2 \otimes I_2,\ \eta H(\mathbf{k})\eta^{-1}=H^{\dagger}(\mathbf{k}).\\
 \eta' &=& \sigma_y \otimes I_2 \otimes I_2, \ \eta' H(k_x,k_y)(\eta')^{-1}=H^{\dagger}(-k_x,k_y),\\
 \eta'' &=& \sigma_x \otimes I_2 \otimes \sigma_z, \ \eta'' H(k_x,k_y)(\eta'')^{-1}=H^{\dagger}(k_x,-k_y). 
\end{eqnarray}
The modified symmetries play a similar role as in the conventional counterpart under the open boundary condition in the $x$ or $y$ direction. For example, the modified symmetry defined by $P'T$ leads to the $PT$ symmetry under the open boundary condition in the $x$ direction because the wavenumber $k_x$ is no longer a good quantum number in this situation. The random real on-site potential and the imaginary noise in the non-Hermitian coupling only satisfy the modified $PT$ symmetry defined by $P'T$ and $P''T$. The exceptional edge modes are robust against these disorders (cf. Fig.~3b in the main text), which implies that the modified $PT$ symmetry acts as the $PT$ symmetry in the effective edge Hamiltonian (cf. Eq.~(3) in the main text) and thus can protect the exceptional edge modes. On the contrary, the real noise in the non-Hermitian coupling considered in Fig.~3c in the main text breaks all the symmetries above and thus opens a gap in the edge bands. We can also check that the imaginary on-site potential considered to calculate the band structures in Fig.~3d in the main text and in Supplementary Fig.~\ref{supplefig6} preserves the conventional $PT$ and $CP$ symmetries and the modified pseudo-Hermiticity. However, this additional term removes the EPs from the edge modes, which is against the naive expectation from the discussion using the symmetry of the effective edge Hamiltonian. We explain the reason in the next paragraph.

The disappearance of exceptional edge modes from the band structures in Fig.~3d in the main text and in Supplementary Fig.~\ref{supplefig6} indicates that the conventional $PT$ and $CP$ symmetries and the modified pseudo-Hermiticity play no roles in protecting the exceptional edge modes. We can explain this from the fact that the operators, $PT$, $CP$, $\eta'$, and $\eta''$, convert an edge mode into another edge mode localized at the opposite side of the system. Thus, the $2\times2$ effective Hamiltonian (cf. Eq.~(3) in the main text) is insufficient to explain what constraint these symmetries impose on the edge modes because it only considers the edge modes localized at one side of the system. To deal with the edge modes at both sides simultaneously, we need a $4\times4$ effective edge Hamiltonian. For both the $2\times2$ and $4\times4$ effective edge Hamiltonian, we can calculate the following topological invariant \cite{Gong2018,Kawabata2019b},
\begin{equation}
\nu = {\rm sgn} (\det H(k_0-\delta) H(k_0 + \delta))
\end{equation}
which guarantees the robustness of EPs in one-dimensional systems. Here, $k_0$ is the wavenumber at the EP, and $\delta$ is a small real number. We confirm that the topological invariant for the $2\times2$ effective edge Hamiltonian is nontrivial, $\nu=-1$, while that for the $4\times4$ effective edge Hamiltonian is trivial, $\nu=1$. Supplementary Figure \ref{supplefig7} shows the change of the eigenvalues around the EP and schematically explains the triviality and nontriviality of the topological invariants. Crossing the EP, the two eigenvalues of exceptional edge modes localized at the same side becomes a pair of complex conjugates from a pair of opposite real numbers or vice versa. On the contrary, the eigenvalues of two edge modes which are transformed into each other by the operators, $PT$, $CP$, $\eta'$, and $\eta''$, take the same real value in a certain range of the wavenumber. Calculating the topological invariant from these eigenvalues, we can confirm its nontriviality (triviality) for the former (latter) case, corresponding to the $2\times2$ ($4\times4$) effective edge Hamiltonian. Therefore, the conventional $PT$ symmetry, the conventional $CP$ symmetry, the modified pseudo-Hermiticity are irrelevant to the protection of exceptional edge modes, while the modified $PT$ symmetry, the modified $CP$ symmetry, the conventional pseudo-Hermiticity, and the chiral symmetry can protect the exceptional edge modes.

Finally, we show the symmetries of other models and that there exist symmetries that can protect the EPs in the edge bands. In the same manner as in the trivial tight-binding model, the nontrivial tight-binding model (cf. Eq.~(2) in the main text and Supplementary Eq.~\eqref{nontrivial-model}) exhibits the $PT$ symmetry, the $CP$ symmetry, the chiral symmetry, and the pseudo-Hermiticity defined as,
\begin{eqnarray}
 PT &=& \sigma_x \otimes \sigma_z,\ PTH(\mathbf{k})(PT)^{-1}=H^{\ast}(\mathbf{k}),\\
 CP &=& \sigma_z \otimes \sigma_y, \ CPH(\mathbf{k})(CP)^{-1}=-H^{\ast}(\mathbf{k}),\\
 \Gamma &=& \sigma_x \otimes \sigma_x, \ \Gamma H(\mathbf{k})\Gamma^{-1}=-H^{\dagger}(\mathbf{k}),\\
 \eta &=& \sigma_z \otimes I_2,\ \eta H(\mathbf{k})\eta^{-1}=H^{\dagger}(\mathbf{k}),
\end{eqnarray}
and the modified $PT$ symmetry, $CP$ symmetry, and pseudo-Hermiticity:
\begin{eqnarray}
 P'T &=& I_2 \otimes \sigma_z, \ P'TH(k_x,k_y)(P'T)^{-1}=H^{\ast}(-k_x,k_y),\\
 P''T &=& \sigma_z \otimes I_2, \ P''TH(k_x,k_y)(P''T)^{-1}=H^{\ast}(k_x,-k_y),\\
 CP' &=& \sigma_y \otimes \sigma_y, \ CP'H(k_x,k_y)(CP')^{-1}=-H^{\ast}(-k_x,k_y),\\
 CP'' &=& \sigma_x \otimes \sigma_x, \ CP''H(k_x,k_y)(CP'')^{-1}=-H^{\ast}(k_x,-k_y),\\
 \eta' &=& \sigma_y \otimes I_2, \ \eta' H(k_x,k_y)(\eta')^{-1}=H^{\dagger}(-k_x,k_y),\\
 \eta'' &=& \sigma_x \otimes \sigma_z, \ \eta'' H(k_x,k_y)(\eta'')^{-1}=H^{\dagger}(k_x,-k_y). 
\end{eqnarray}
Again, the conventional $PT$ and $CP$ symmetry, and the modified pseudo-Hermiticity play no roles in protecting exceptional edge modes, which we can confirm from the result in Supplementary Fig.~\ref{supplefig2}d. The topological laser analyzed in Fig.~4 and 5 in the main text has only the following modified $PT$ symmetry, 
\begin{eqnarray}
 P'T &=& I_2 \otimes \sigma_z, \ P'TH(k_x,k_y)(P'T)^{-1}=H^{\ast}(-k_x,k_y),\\
 P''T &=& \sigma_z \otimes I_2, \ P''TH(k_x,k_y)(P''T)^{-1}=H^{\ast}(k_x,-k_y),
\end{eqnarray}
as symmetries related to the protection of exceptional edge modes. The continuous model analyzed in Fig.~6 in the main text is rewritten as
\begin{eqnarray}
 H(\mathbf{k}) &=& [M+\beta(k_x^2+k_y^2)]I_2 \otimes \sigma_z + a(k_x^2-k_y^2) I_2\otimes\sigma_x \nonumber\\
 &{}& + 2ak_x k_y \sigma_z \otimes \sigma_y + i\frac{b+b'}{2}\sigma_x \otimes \sigma_x + \frac{-b+b'}{2}\sigma_x \otimes \sigma_y.
\end{eqnarray}
We can confirm that this model exhibits the modified $PT$ symmetry
and the chiral symmetry defined as,
\begin{eqnarray}
 P'T &=& \sigma_z \otimes I_2, \ P'TH(k_x,k_y)(P'T)^{-1}=H^{\ast}(-k_x,k_y),\\
 P''T &=& \sigma_z \otimes I_2, \ P''TH(k_x,k_y)(P''T)^{-1}=H^{\ast}(k_x,-k_y),\\
 \Gamma &=& \sigma_y \otimes \sigma_y, \ \Gamma H(\mathbf{k})\Gamma^{-1}=-H^{\dagger}(\mathbf{k}).
\end{eqnarray}
Finally, the effective Hamiltonian of the chiral active matter model only shows the chiral symmetry
\begin{eqnarray}
 \Gamma = \sigma_x \otimes I_3, \ \Gamma H(\mathbf{k})\Gamma^{-1}=-H^{\dagger}(\mathbf{k}),
\end{eqnarray}
where $I_3$ is a $3\times3$ identity matrix.

\begin{figure}[t]
\includegraphics[width=80mm,bb=0 0 400 600,clip]{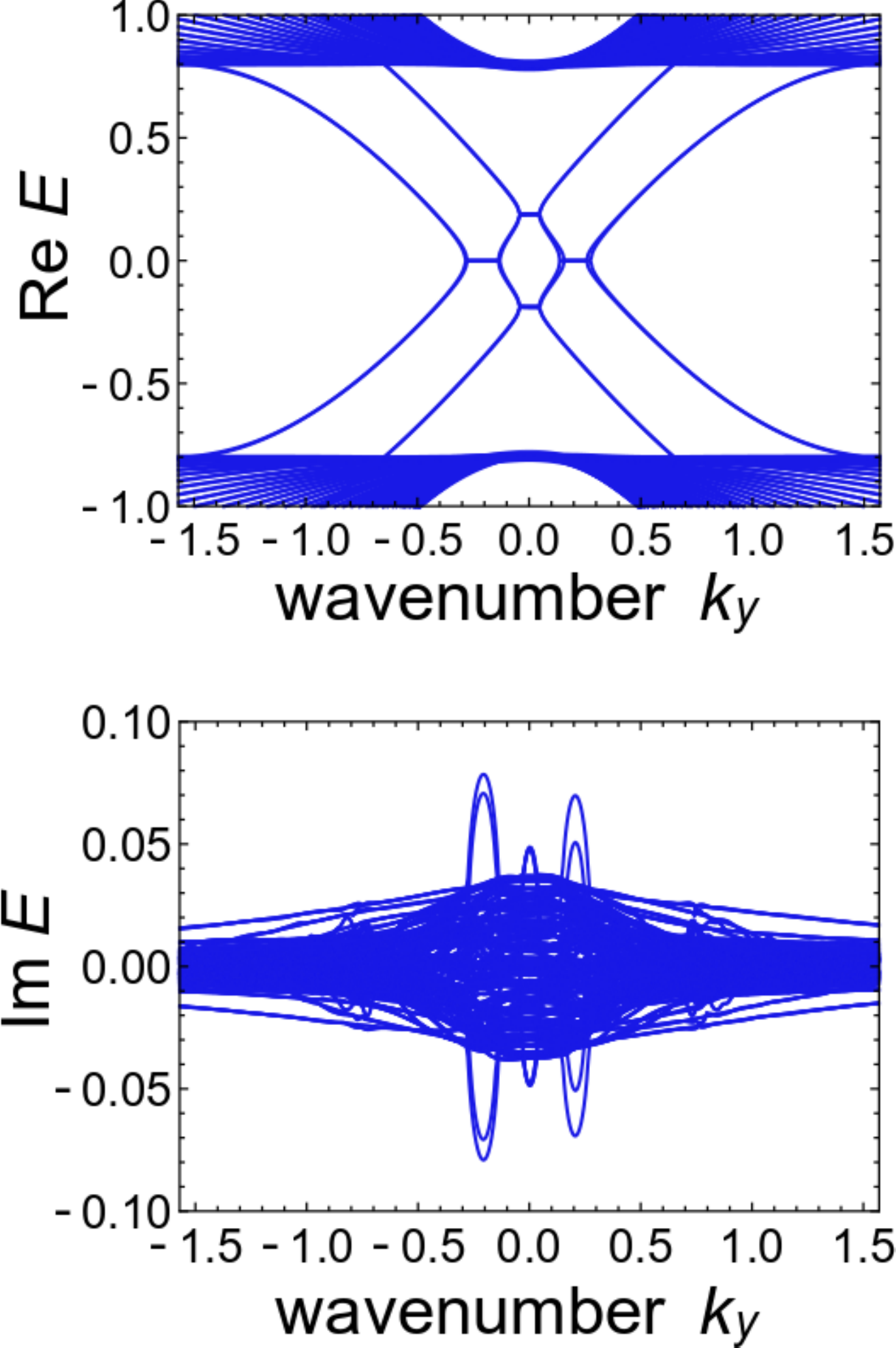}
\caption{\label{supplefig8}
Exceptional edge modes under chiral-symmetry-preserving disorder. We arrange 50 sites in the $x$ direction with the open boundary and impose the periodic boundary condition in the $y$ direction. The parameters used are $u=-1$, $c=0.2$, $\beta=0.14$, $\beta'=0.06$ and $\gamma=0.05$. The noise width is $w=0.05$, ($W=0.05$) for the disorder $id(x)\sigma_x\otimes I\times I$ (the imaginary noise in the non-Hermitian coupling).
}
\end{figure}

\subsection{Exceptional edge modes protected by the chiral symmetry}Since EPs can be protected by the chiral symmetry and the topology of a system, exceptional edge modes can also exist robustly under the chiral symmetry. To confirm the robustness of exceptional edge modes against chiral-symmetry-preserving disorder, we consider the two-layered non-Hermitian Bernevig-Hughes-Zhang model with the imaginary noise in the non-Hermitian coupling considered in Fig.~3 in the main text and disorder $id(x)\sigma_x\otimes I\otimes I$. Here, $d(x)$ takes a random real value for each $x$ from uniform distributions ranging $[-w,w]\in\mathbb{R}$. We use $w=0.05$, and the other parameters are the same as used in Fig.~3 in the main text. These disorders break the conventional and modified $PT$, $CP$ symmetries, and pseudo-Hermiticity. Therefore, the chiral symmetry, $\Gamma = \sigma_x \otimes \sigma_z \otimes \sigma_x$, $\Gamma H(\mathbf{k})\Gamma^{-1}=-H^{\dagger}(\mathbf{k})$, is the only symmetry that can protect the exceptional edge modes in this disordered system. Supplementary Figure \ref{supplefig8} shows the edge band structure under the disorders. The EPs remain and protect the existence of edge modes, which implies that the chiral symmetry can also support the robustness of exceptional edge modes. We note that the eigenvalues do not appear as real values or pairs of complex conjugates, while the $PT$ symmetry and the pseudo-Hermiticity lead to such constraint on the band structure. Since the chiral symmetry is the sole symmetry that can protect the exceptional edge modes in the active matter model analyzed in Fig.~7 in the main text as discussed in the previous section, this result also guarantees the robustness of exceptional edge modes there.

We can also confirm that this model is topologically trivial in the conventional sense and thus clearly breaks the bulk-edge correspondence. The model belongs to AIII class, which only has the chiral symmetry and thus can only exhibit a trivial invariant for a real line gap \cite{Zhou2019,Kawabata2019}. The topological invariant for an imaginary line gap cannot be defined because the model has no imaginary line gaps. The topological invariant for a point gap in AIII class systems can be defined by the Chern number of $iH\Gamma$ \cite{Kawabata2019b} (note that $iH\Gamma$ is Hermitian and if it has a zero-energy eigenstate $\psi$, $\Gamma\psi$ is a zero-energy eigenstate of the non-Hermitian Hamiltonian $H$). We numerically confirm the triviality of the topological invariant for the present model with constant $d(x)=0.05$ and without the real noise in the non-Hermitian coupling by using the numerical methods proposed by Fukui, T., Hatsugai, Y., and Suzuki, H. \cite{Fukui2005}. Here we set the other parameters to be the same as in Supplementary Fig.~\ref{supplefig8}. Therefore, the model analyzed here is topologically trivial in the conventional sense, while it exhibits robust exceptional edge modes. Since the model exhibits no spatial symmetries like the $PT$ symmetry, the bulk-edge correspondence should indicate the consistency between the topological invariant predicted from the periodic table \cite{Gong2018,Zhou2019,Kawabata2019} and the existence of the robust edge modes, which is broken in the present model.

\end{document}